\documentclass[utf8]{frontiersSCNS} 

\usepackage{url,hyperref,lineno,microtype,subcaption}
\usepackage[onehalfspacing]{setspace}

\usepackage{graphicx}
\usepackage{amsmath,epsfig,rotating,amssymb}

\newcommand\bb[1] {   \mbox{\boldmath{$#1$}}  }

\newcommand\bcdot{\bb{\cdot}}
\newcommand\btimes{\bb{\times}}

\newcommand{ \PD }[2]{ \frac{ \partial #1 }{ \partial #2 } }

\def\keyFont{\fontsize{8}{11}\helveticabold }
\def\firstAuthorLast{Hennebelle {et~al.}} 
\def\Authors{Patrick Hennebelle\,$^{1,*}$ and Shu-ichiro Inutsuka\,$^{2}$}
\def\Address{$^{1}$ Laboratoire AIM, Paris-Saclay, CEA/IRFU/SAp -- CNRS --
      Universit\'e Paris Diderot, 91191 Gif-sur-Yvette Cedex, France
      LERMA (UMR CNRS 8112), Ecole Normale Sup\'erieure, 75231 Paris Cedex,
      France 
$^{2}$ Department of Physics, Nagoya University, Chikusa-ku, Nagoya 464-8602, Japan }
\def\corrAuthor{Patrick Hennebelle}

\def\corrEmail{patrick.hennebelle@cea.fr}

\begin{document}
\onecolumn
\firstpage{1}

\title[Magnetic field in molecular clouds]{The role of magnetic field in molecular cloud formation and evolution} 

\author[\firstAuthorLast ]{\Authors} 
\address{} 
\correspondance{} 

\extraAuth{}

\maketitle

\begin{abstract}

\section{}
We review the role that magnetic field may have on the formation and evolution of molecular clouds. 
After a brief presentation and main assumptions leading to ideal MHD equations, their most 
important correction, namely the ion-neutral drift is described. The nature of the multi-phase interstellar medium
(ISM) and the thermal processes that allows this gas to become denser are presented. Then we discuss our 
current knowledge of compressible magnetized turbulence, thought to play a fundamental role in the ISM. We also 
describe what is known regarding the correlation between the magnetic and the density fields.
Then the influence that magnetic field may have on the interstellar filaments and  the molecular clouds is 
discussed, notably the role it may have on the prestellar dense cores as well as regarding the formation 
of stellar clusters. Finally we briefly review its possible effects on the formation of molecular clouds 
themselves. We argue that  given the magnetic intensities that have been measured, it is likely that 
magnetic field is 
i) responsible of reducing the star formation rate in dense molecular cloud gas by a factor of a few, ii) strongly shaping the 
interstellar gas by generating a lot of filaments and reducing the numbers of clumps, cores and 
stars, although its exact influence remains to be better understood. 
Moreover at small scales,  magnetic braking 
is likely a dominant process that strongly modifies the outcome of the star formation process.  
Finally, we stress that by inducing the formation of  more massive stars,  magnetic field could possibly 
enhance the impact of stellar feedback.

\tiny
 \keyFont{ \section{Keywords:} magnetic field, molecular clouds, star formation, gravity, turbulence } 
\end{abstract}

\section{Introduction}
The interstellar cycle, which takes place within galaxies, is  fundamental for our universe
as it controls the formation of stars and therefore the evolution of galaxies. Yet given 
the broad range of spatial scales and the profusion of physical processes involved, 
 our understanding is still very incomplete. Amongst other processes, namely gravity, compressible turbulence, 
radiation, cosmic rays and stellar feedback, magnetic field is also contributing significantly to the 
evolution of the interstellar medium (ISM) and more specifically to the formation of stars. 
As a matter of evidence, the magnetic energy in the ISM is comparable to the other energies as 
for example the kinematic one. Deciphering the various roles that magnetic field is playing is 
however not obvious, i) because measuring it remains a challenge, ii) because 
magnetic field is not a mere pressure and is highly non-isotropic in nature, iii) because observations 
do not allow us to easily vary the parameters as it is possible to do in experiments.
This however can be done in numerical simulations where the influence of a specific 
parameter, like the magnetic intensity, can be modified and studied.  

This review is dedicated to the role magnetic field is playing in the formation and evolution of 
molecular clouds. Given the complex multi-scale nature of these latter, this represents 
a challenge as several physical processes and astrophysical objects have to be discussed, 
in particular because as stressed above, the magnetic field is strongly interacting with 
other phenomena, that need to be described for self-consistency.

The plan of the paper is as follows. In the second section, we describe the equations 
of the magneto-hydrodynamics (MHD) that are used to 
compute and predict the evolution of molecular clouds. We give some ideas of 
how these equations are inferred first in the ideal MHD framework, that 
is to say when the fluid and the magnetic field are perfectly coupled. Then 
we briefly discuss the most important correction that must be taken into 
account in molecular clouds, namely the ion-neutral drift or  ambipolar 
diffusion. 
In the third section, the multi-phase nature of the ISM is discussed: 
how the gas cools and heats, the principle of thermal instability and its
non-linear regime. The role that magnetic field may have in the 
transition from warm atomic gas into cold and dense gas is emphasized. 
In the fourth section, the nature of the turbulence in the ISM 
is presented. First some elements of the magnetized incompressible turbulence 
are given, stressing the ideas and problems  that are still debated. 
Second the more realistic compressible and multi-phase magnetized turbulence
is addressed, reporting the various numerical studies that have been performed. 
The influence that the ion-neutral drift may have on turbulence is discussed. 
Section five is specifically dedicated to the correlation between density and magnetic 
field including the magnetic intensity and the magnetic orientation. 
Section six is  specifically dedicated to puzzling astrophysical objects, 
namely the filaments that, in a sense, constitute sub-structures of molecular clouds. 
The question of their formation, the physical origin of the possible characteristic width 
that has been recently inferred and their fragmentation in star forming cores are discussed.
In section seven  the molecular clouds themselves are eventually addressed. We start 
by reviewing the role that ambipolar diffusion may have in the magnetically dominated clouds, 
particularly regarding the fundamental question of the low efficient formation of stars 
in galaxies. Then the properties of the prestellar cores which form in dedicated 
numerical simulations of these clouds are described stressing the effect that magnetic field 
may have. Finally the role that magnetic field may have in stellar clusters 
formation is presented. 
In section eight we briefly review the works in which the impact of the magnetic field 
on molecular cloud formation has been addressed. 
Section nine concludes the paper.

\section{MHD equations}
For the sake of completeness and because readers may find it 
useful, a short derivation and discussion of the 
MHD equations is given. 
We start with the ideal MHD, which amongst other approximations
assume the  non-relativistic limit, that is to say the  fluid velocities 
are much smaller than the speed of light $c$.
 We also discuss
a correction to these set of equations in the presence of  
  neutral particles since the ionization degree in the ISM is small. 

\subsection{Ideal MHD}
The equations of ideal MHD assume that the fluids are perfect
 conductors.   The Lorentz force, which is the force that
the electromagnetic fields $\bb{E}$ and $\bb{B}$ exert on the
fluid must be taken into account.
 The electromagnetic fields evolution is obviously described by the
Maxwell equations. Written in CGS  units, these equations are
\begin{eqnarray}
\bb{\nabla} \bcdot \bb{B} &=& 0 \label{solenoidal_eq}, \\
\bb{\nabla} \bcdot \bb{E} &=& 4 \pi \rho_e,  \\
c \bb{\nabla} \btimes \bb{E} &=& -\frac{\partial \bb{B}}{\partial t}, \label{faraday_eq} \\
c \bb{\nabla} \btimes \bb{B} &=& 4 \pi \bb{j} + \frac{\partial \bb{E}}{\partial t}, \label{ampere_eq} 
\end{eqnarray}
 $\rho_e$ and $\bb{j}$ are the fluid charge and  current
densities. The equation for charge
conservation links these two quantities 
\begin{equation}
\frac{\partial \rho_e}{\partial t} + \bb{\nabla} \bcdot \bb{j} = 0 \, .
\end{equation}

While in a perfect conductor at rest, $\bb{E_R}$  vanishes, 
the situation is different when it moves.
 The rest fields $\bb{E_R}$ and $\bb{B_R}$ and the fields
in the observer frame,
$\bb{E}$ and $\bb{B}$ 
are related using  the Lorentz transformation, 
as detailed by e.g. \citet{landau1960}, \citet{shu1992},  \citet{spruit2013}. 
Considering
the Lorentz force $\bb{F}$ and $\bb{F_R}$, we get:
\begin{eqnarray}
      \bb{F} &=& q (\bb{E} + \frac{\bb{v}}{c} \btimes \bb{B}) \, , \\
\bb{F_R} &=& q\bb{E_R} \, .
\end{eqnarray}
Since the force does not depend on the reference frame:
$\bb{F}=\bb{F_R}$ and therefore 
\begin{eqnarray}
      \bb{E_R} &=& \bb{E} + \frac{\bb{v}}{c} \btimes \bb{B} \, , \\
\bb{B_R} &=& \bb{B} \, .
\end{eqnarray}
Since the perfect conductor assumption is made, $\bb{E_R}=\bb{0}$, and 
\begin{equation}
      \bb{E} =  - \frac{\bb{v}}{c} \btimes \bb{B} \, .
\label{E_field_eq}
\end{equation}
Combining it with Eq.~(\ref{faraday_eq}) we obtain:
\begin{equation}
\frac{\partial \bb{B}}{\partial t} = \bb{\nabla} \btimes \left( \bb{v}
\btimes \bb{B} \right).
\end{equation}

\noindent
The Lorentz force per unit volume,
$\bb{f_L}$,  can be expressed as
\begin{equation}
\bb{f_L}=\rho_e \bb{E} + {1 \over c} \bb{j} \btimes \bb{B} \, .
\end{equation}
 In the non-relativistic limit, the
displacement current  in Eq.~(\ref{ampere_eq}) can be  neglected, leading to
\begin{equation}
  \bb{\nabla} \btimes \bb{B} = {4 \pi \over c} \bb{j} \, .
\label{j_eq}
\end{equation}
Since local electroneutrality is assumed, we have
$\rho _E = 0$ and 
\begin{equation}
\bb{f_L}=\frac{(\bb{\nabla} \btimes \bb{B})\btimes \bb{B}}{4 \pi} \, .
\end{equation}
This leads to the standard form of the ideal  MHD equations 
\begin{eqnarray}
\frac{\partial \rho}{\partial t} + \bb{\nabla} \bcdot (\rho \bb{v})
&=& 0, \\
\rho \left[ \frac{\partial \bb{v}}{\partial t} + (\bb{v} \bcdot
  \bb{\nabla}) \bb{v} \right] &=& -\bb{\nabla} P +
\frac{(\bb{\nabla} \btimes \bb{B}) \btimes \bb{B}}{4\pi} \label{impulsion_eq}, \\ 
\rho \left[ \frac{\partial \bb{e}}{\partial t} + (\bb{v} \bcdot
  \bb{\nabla}) \bb{e} \right] &=& - P (\bb{\nabla} \bcdot
\bb{v}) \label{internal_nrj_eq} - \rho {\mathcal L},  \\
\frac{\partial \bb{B}}{\partial t} &=& \bb{\nabla} \btimes (\bb{v}
\btimes \bb{B}). \label{induction_eq}
\end{eqnarray}
where ${\mathcal L}$ is the net loss function and describes the radiative heating and cooling of the gas.
This must be complemented by an  equation of state  to close the system of
equations. A perfect gas is a good assumption for the ISM,  $P=(\gamma -
1) \rho \epsilon$, where $\gamma$ is the adiabatic index of the gas. 

\noindent

It is useful to get a better insight and physical interpretation of the MHD 
equations to rewrite the Lorentz force as
\begin{equation}
\bb{f_L}=\frac{(\bb{\nabla} \btimes \bb{B}) \btimes
  \bb{B}}{4\pi}=-\bb{\nabla}\left( \frac{B^2}{8\pi} \right) + \frac{(\bb{B}
\bcdot \bb{\nabla}) \bb{B} }{4\pi}
\end{equation}
The first term  is called 
the magnetic pressure. The second 
is the magnetic tension
(see the detailed discussion page 13 in the
lecture by \citet{spruit2013}).

\subsection{Non-ideal MHD: the ion-neutral drift}
\label{ambi}
In many situations,  ideal MHD is not a sufficiently good assumption and 
additional effects need to be accounted for. In the context of molecular clouds
the dominant correction is the so-called ambipolar diffusion. 
Since the neutrals are not charged they are not subject to the 
Lorentz force which applies  only on the ions. However 
through collisions the neutrals and the ions exchange momentum
and therefore the Lorentz force has an influence on the neutrals through 
the ions. If the number of ions is large, i.e. if the 
ionisation is high, the number of collisions is expected to be large 
and ideal MHD remains a good approximation.  
However in molecular clouds the ionisation is 
usually of the order of $10^{-7}$ and therefore the two fluid 
are not perfectly coupled. The ions drag the field lines and 
drift with respect to the neutrals implying that the latter can 
cross the field lines. The field is not frozen in the gas anymore. 
Because of the low ionisation, it is thus possible to
neglect the inertia of the ions and a reasonable  assumption is that
of the equilibrium between the Lorentz force and the drag force
exerted on the ions. This leads to:
\begin{equation}
{(\nabla \times {\bf B}) \times {\bf B} \over 4 \pi} = 
\gamma _{ad} \rho \rho_i ( {\bf V}_i - {\bf V}),
\label{coupling}
\end{equation}
where $\rho_i$ and {\bf$V_i$} are the ion density and velocity
respectively, $\gamma _{ad} \simeq 3.5 \times 10^{13}$ cm$^3$ g$^{-1}$
s$^{-1}$ is the drag coefficient \citep{mouschovias1981}.
Eq.~(\ref{coupling}), gives the ion velocity as a 
function of the neutral velocity and the Lorentz force. 
Combining it with  the induction equation
one gets
\begin{equation}
\partial _t {\bf B}  + \nabla \times  ({\bf B} \times {\bf V}) = 
\nabla \times \left( {1 \over 4 \pi \gamma _{ad} \rho \rho_i}  ( (\nabla \times {\bf B}) \times {\bf B}) \times {\bf B} 
 \right). 
\label{induc}
\end{equation}
The left-hand side is the induction equation of ideal MHD. The right-hand side 
 describes  the  ion-neutral drift.
It is not rigorously speaking a diffusion term although 
it entails second order spatial derivatives. 
From this equation a typical time scale 
for  ambipolar diffusion can  easily be  inferred
\begin{eqnarray}
\tau_{\rm ad} \simeq {4 \pi \gamma _{ad} \rho \rho_i L ^2 \over B^2 },  
\label{time}
\end{eqnarray}
where $L$ is the characteristic  spatial scale of  the problem, which 
could be the size of the prestellar cores or the filaments as described below. Ionization equilibrium 
leads to  $\rho_i = C \sqrt{\rho}$, where 
$C=3 \times 10^{-16}$ cm$^{-3/2}$ g$^{1/2}$. 

As Eq.~(\ref{coupling}) neglects the ion inertia, it is called 
the strong coupling limit \citep[e.g.][]{shu1992,maclow1995,jacques2012}. 
Ideally, it is necessary to consider two fluids 
the neutral and the ions coupled through the collisional term. 
The difficulty however with this approach is that the Alfv\'en speed 
associated to the ions is several orders of magnitude larger than 
the Alfv\'en speed associated to the neutrals. This makes numerical 
simulations very difficult to perform because the timesteps are then 
very small. For this reason an alternative approximation, called the 
heavy ion approximation has been developed \citep{limckee2006}. It consists in artificially 
increasing the mass of the ions 
to lower their Alfv\'en speed while modifying the ion-neutral cross-section to maintain 
constant the friction coefficient.

Finally, let us mention that the ion-neutral friction leads to energy dissipation and therefore
constitute a source of heating in Eq.~(\ref{internal_nrj_eq}) which is equal 
to 
$\gamma _{ad} \rho \rho_i ( {\bf V}_i - {\bf V})^2$.


\section{The formation of dense gas in the ISM}
Here we describe how the formation of dense gas out of diffuse atomic gas is achieved 
in the ISM. A brief description of the cooling and heating processes, essential to 
understand how the ISM becomes denser is given. We then describe the principle of 
thermal instability on the role magnetic field may have. Finally, a dynamical scenario 
for the formation of molecular clouds is sketched, stressing how magnetic field 
is acting.

\subsection{Thermal Structure of ISM and Thermal Instability}
\label{Sec:Thermal}
\begin{figure}[h]
  \includegraphics[width=8.7cm]{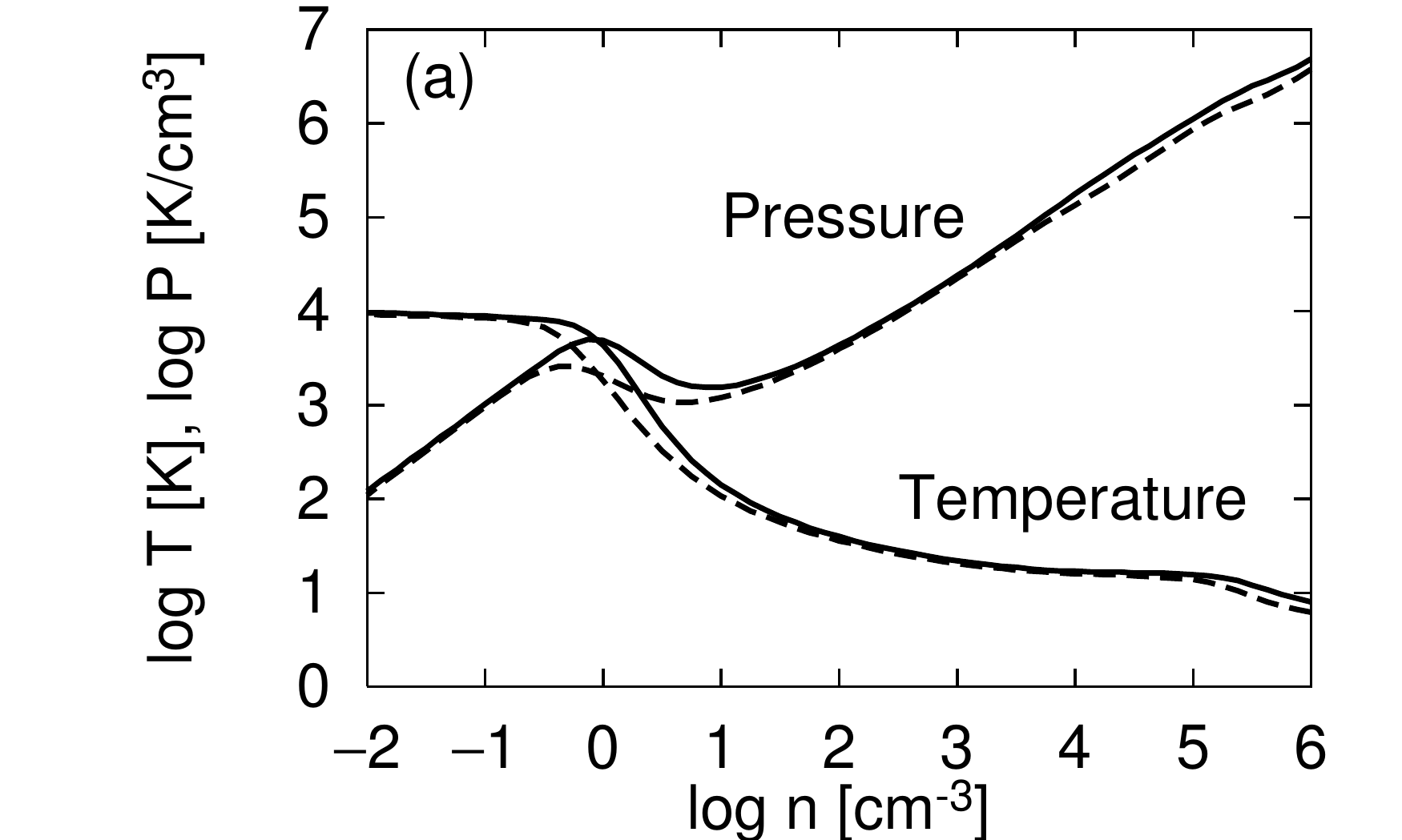}
  \includegraphics[width=8.7cm]{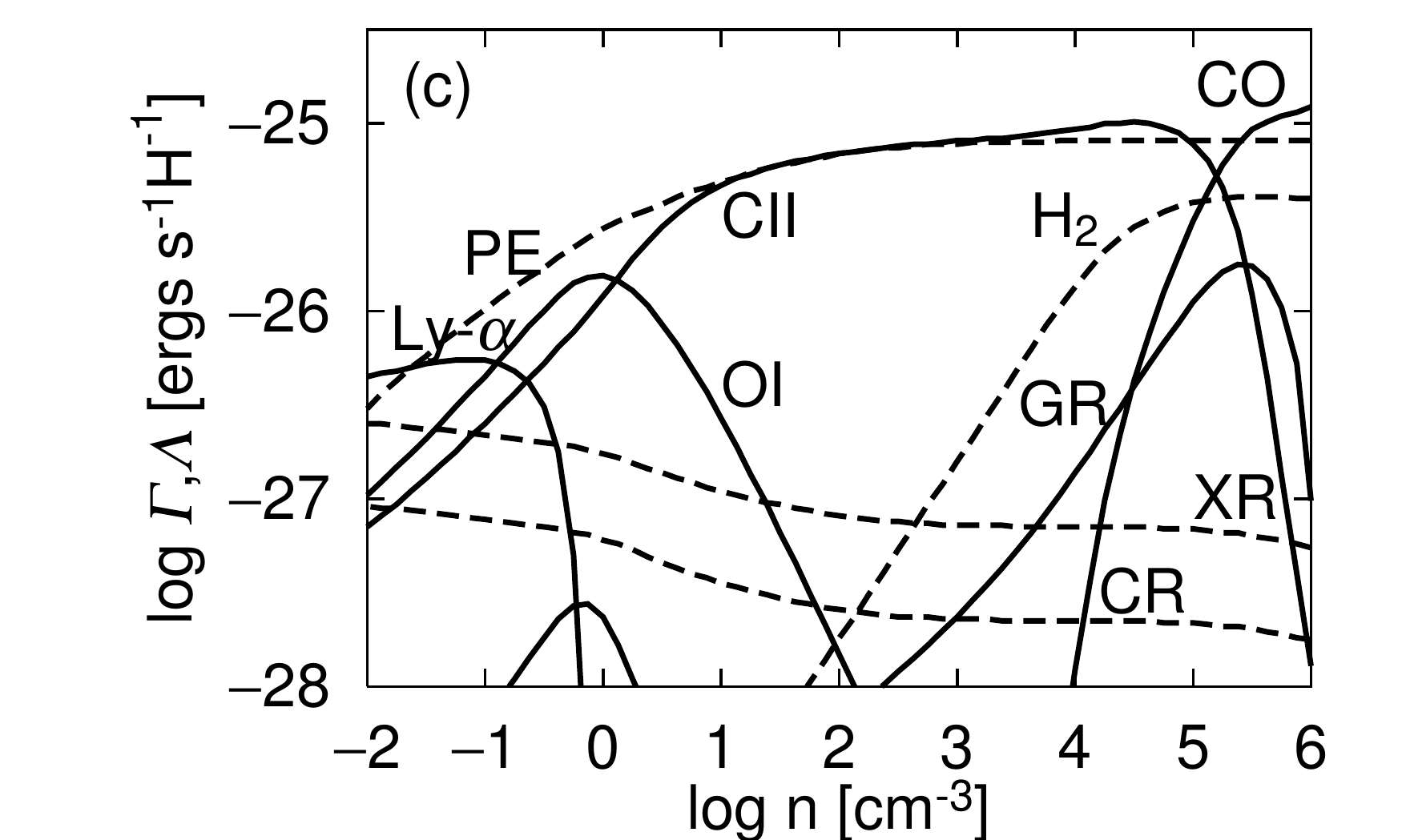}
\caption{A phase diagram of ISM and important heating and cooling processes shown in \citet{koyama2000}. 
  (a) Equilibrium temperature and pressure
with absorbing column density of  10$^{19}\, {\rm cm^{-2}}$ (solid lines),
10$^{20}\, {\rm cm^{-2}}$ (dash lines). 
  (c) Heating (dashed lines) and cooling rates (solid lines) for unshielded gas per hydrogen
 nucleus at equilibrium.
Heating processes are,
photoelectric effect from small grains and PAHs (PE),
X-ray (XR),
Cosmic-ray (CR), and
H$_2$ formation/destruction.
Cooling processes are
CII fine-structure (CII),
OI fine-structure (OI), 
Hydrogen Lyman-$\alpha$ (Ly-$\alpha$),
CO rotation/vibration line (CO), and
atomic and molecular collisions with dust grains (GR).
Reproduced from \citet{koyama2000} with permission of ApJ.}  
\label{fig:equilibrium}
\end{figure}

In this section, our knowledge about thermal physics of ISM is presented.
To calculate the equilibrium temperature of ISM as a function of gas density,
one must equate  heating and cooling function taking into account several
 physical processes. 
The detailed analysis for the thermal equilibrium state in the neutral atomic phase can be found, for example, in \citet{wolfire2003} 
while \citet{koyama2000}
 \citep[see also e.g.][]{glover2012,gong2017} extended their calculation for unshielded gas 
up to gas densities
$10^3 \, {\rm cm^{-3}}<n<10^6 \, {\rm cm^{-3}}$
(see Figure \ref{fig:equilibrium}).
The main heating mechanisms are the photoelectric emission from small grains and PAHs, ionization 
by cosmic rays and soft X-rays, and the formation and photodissociation of H$_2$.
The local FUV field is supposed to be on the order of Habing's value ($G_0=1.7$ and is adopted in Figure \ref{fig:equilibrium}).
The dominant cooling  processes are 
 the line emission from H, C, O, Si, and Fe, by rovibrational lines from H$_2$ and CO, as well 
as  by atomic and molecular collisions with dust grains.
The transition to the molecule-dominated phase depends on both gas density and column density as well 
as the radiation field (see \citet{sternberg2014} for analysis and discussion of the relative 
importance of H$_2$ self-shielding and dust shielding).  The chemistry and cooling in gas with a
 range of density, column, metallicity, and radiation fields is discussed e.g. in  \citet{glover2012}
 and \citet{gong2017}. 
To describe these thermal processes 
a set of three time-dependent equations for ionization and recombination 
of hydrogen, and formation and dissociation of molecules should be solved.
Self-shielding effects must also be taken into account to calculate the 
 H$_2$ photodissociation.
Figure \ref{fig:equilibrium} portrays the resulting 
 temperature, pressure, and the relevant chemical species 
as functions of number density for unshielded gas. At high column density, inside molecular clouds, the dominant molecular cooling process is due to the 
CO molecules (at  densities above 10$^5$ cm$^{-3}$ dust cooling becomes dominant) and the heating one comes 
from cosmic rays. 
A complete thermal balance description of the high density gas, which can be found for instance 
  in \citet{neufeld1995}, is beyond the scope of the present 
review.

\begin{figure}[h]
  \includegraphics[width=8cm]{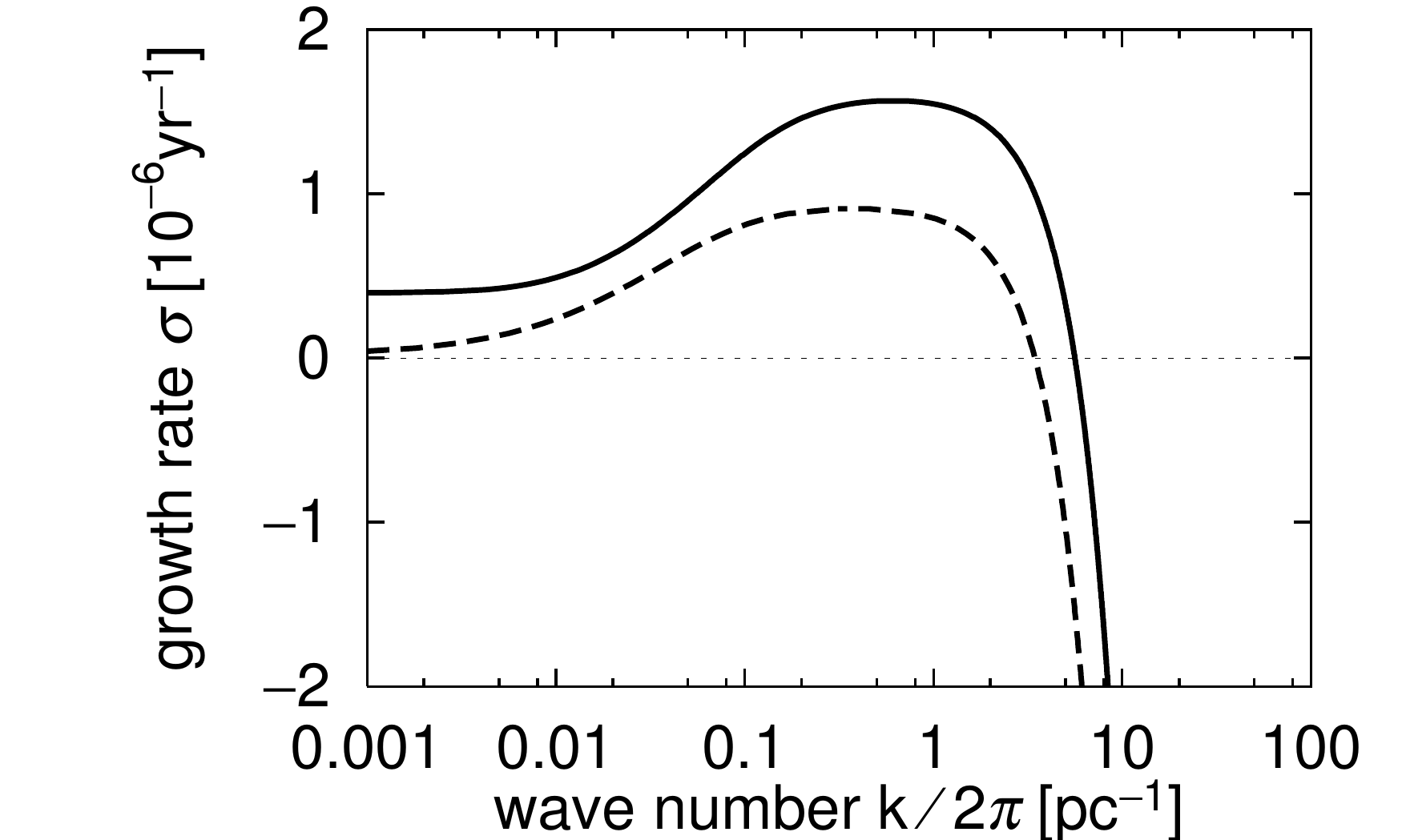}
  \caption{ The dispersion relation for condensation mode 
            of thermal instability analyzed in \citet{koyama2000}. 
            The dashed curve denotes the classical result 
            for the case of thermal equilibrium unperturbed state. 
            The solid curve denotes for the case of an isobarically 
            contracting unperturbed state. 
            Reproduced from \citet{koyama2000} with permission of ApJ.}  
            \label{Fig:DR}
\end{figure}

The basic property of thermal stability can be related to the slope of heat-loss function, $\mathcal{L}=\rho \Lambda-\Gamma$, where $\rho \Lambda$ is the cooling function per volume and $\Gamma$ is the heating function. 
\citet{field1965} studied in details the stability conditions of a uniform 
medium subject to heating and cooling. In particular, he inferred 
the isobaric criterion which is  given by 
\begin{equation}
       \left( \PD{\mathcal{L}}{T} \right)_{\rm P} < 0 \Leftrightarrow 
\left( { \partial P \over \partial \rho} \right) _{\mathcal{L}} < 0.
\end{equation}
Mathematically, it corresponds to $p<1$ in the case $\Lambda \propto T^p$. 
Typically this unstable phase occurs for temperature between 
$\sim$ 100 and $\sim$ 5000K.
Figure \ref{Fig:DR} shows the growth rate of thermal instability as a function of the wavelength of the linear perturbation. 
The dashed curve corresponds to the case of the unperturbed state in equilibrium. 
The thermal conduction tends to make the system isothermal, and hence, it stabilizes the perturbations with sufficiently small wavelengths. 
The critical wavelength (the largest wavelength stabilized by thermal conduction) is called ``Field length'' named after the pioneer of this analysis:  
\begin{equation}
      \lambda_{\rm c} 
      = 2\pi \left\{ \frac{\rho}{K} \left[ \PD{\mathcal{L}}{T} \right]_{\rm P} \right\}^{-1/2} 
   \sim \sqrt{\frac{KT}{\rho^2 \Lambda}} \equiv \lambda_{\rm F} , 
\end{equation}
where $K$ denotes the coefficient of thermal conduction. 

In the case where the spatially uniform perturbed state is not in thermal 
balance, the criterion for the instability is $p<2$ for $\Lambda \propto T^p$, and hence, the range of the unstable temperature becomes wider \citep{schwarz1972}.  
 The dispersion relation for isobarically 
cooling medium is  portrayed by the solid curve in Figure \ref{Fig:DR}
 \citep{koyama2000}.
 The growth rate presents  a peak at a wavelength that is about ten times
 larger than the Field length implying that 
 thermal instability tends to produce structures  larger than the Field length.

\subsection{The Effect of Magnetic Field on Thermal Instability}
The effect of magnetic field on the linear growth of thermal instability was studied in detail by \citet{ames1973}. 
Obviously a sufficiently strong magnetic field suppresses the motion perpendicular to the magnetic field lines. 
This is because in slab geometry, the magnetic pressure is simply proportional to the density square (as 
magnetic field is proportional to  density), therefore the increase of the magnetic pressure 
can compensate for the decrease of the thermal pressure.
However the perturbations in the direction along the magnetic field are not suppressed and remain unstable, if the cooling function
 satisfies the instability criteria. 
The non-linear development of the thermal instability has been studied by various authors 
\citep{henne2000,piontek2004,inoue2007,vanloo2007,inoue2008,inoue2009,choi2012}
while the effects of non-ideal MHD  on  thermal instability have been  studied by various papers.
\citet{inoue2007} have done one-dimensional two-fluid simulations where neutral and charged components are 
self-consistently described as two fluids.  They found that 
regardless of the initial conditions used to set up the simulation, the magnetic field strength in dense regions 
ends up being a few $\mu$G.

\subsection{Formation of Molecular Clouds}
\begin{figure}[h]
\includegraphics[width=17cm]{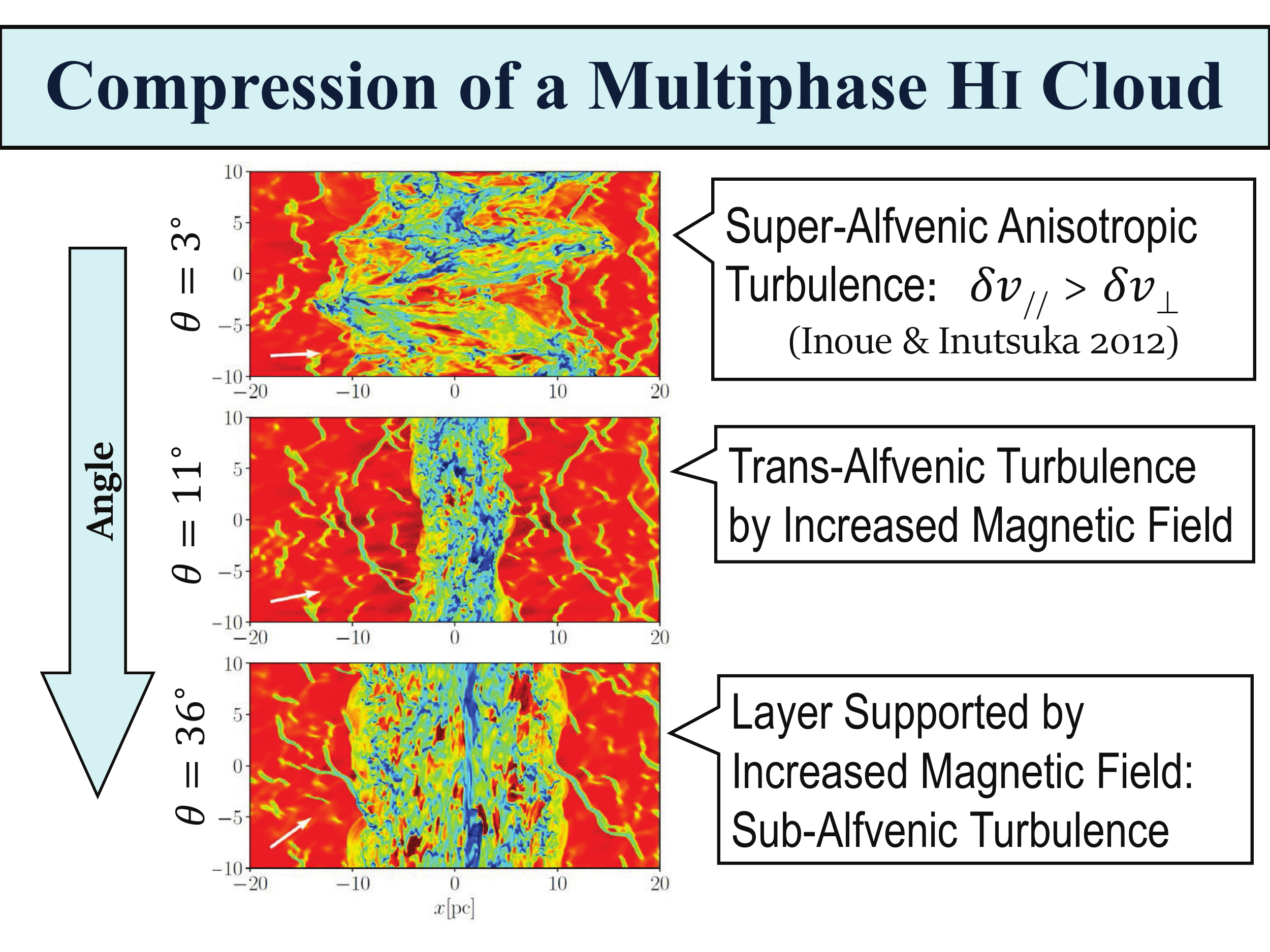}
  \caption{ 
      The result of compression of multiphase HI clouds by shock waves  \citep{iwasaki2018}. 
The column density is shown (red stands for WNM while Blue-green represents CNM).
      The relative angle ($\theta$) between the shock wave propagation direction and the mean 
magnetic field is 3 degrees
 (upper panel), 11 degrees (middle panel) and 36 degrees (lower panel), respectively.}
            \label{Fig:MagLayer}
\end{figure}
It is believed that most of the volume in the thin ($\sim 10^2$pc) disk of our Milky Way Galaxy is filled by 
warm neutral medium (WNM) and warm or hot ionized medium \citep[e.g.][]{ferriere2001}.  
In contrast, a significant fraction of gaseous mass resides in the cold dense medium that occupies only a small 
fraction of the spatial volume. 
Sufficiently dense ($>10^3 {\rm cm}^{-3}$) gas tends to be in large ( $>$ a few pc ) clouds whose column 
density is sufficiently large to protect molecular hydrogen against external 
dissociating radiation. 
Giant molecular clouds are observed to be the sites of massive star formation. 
Once a massive ($>10 M_{\odot}$) star is created in a molecular cloud, it radiates intense ultraviolet radiation inside the cloud that results in the creation of an expanding ionized region, i.e., HII region.  
The expanding HII regions are  supposed not only to quench the further star formation, 
but also to destroy the parental molecular clouds \citep{dale2011,walch2012,dale2013,geen2015,geen2017,gavagnin2017,kim2018}.  
Since we suppose that our Milky Way Galaxy is in a steady state over a timescale of Galactic 
rotation ($\sim 10^8$yr), 
molecular clouds are continuously created at a rate that compensate the destruction by massive stars. 

How are the cold dense clouds created? 
According to the phase diagram of ISM shown in Section \ref{Sec:Thermal}, we can identify that the formation 
process of cold dense HI gas ($>10 {\rm cm}^{-3}$) from WNM
 should be a phase transition dynamics that 
increase the density abruptly  \citep{henne1999,koyama2000}. 
Several studies
\citep{koyama2002,piontek2004,audit2005,heitsch2006,vazquez2006} have shown that this phase transition always results in the creation of long-lasting turbulent motions where cold HI clumps are embedded in WNM.  
The amplitude of turbulent velocities of cold HI gas tends to be a few km  s$^{-1}$, a fraction of the sound speed of the WNM (10 km s$^{-1}$). 
Therefore, the resulting turbulence appears to be supersonic with respect to the sound speed of cold medium ($\sim 1$km s$^{-1}$) but actually 
subsonic with respect to the WNM, which constitutes the inter clump medium. Therefore the turbulence of the atomic interstellar gas 
is a combination of subsonic and supersonic motions. Let us stress that the CNM tends to behave as a supersonic gas even so it is 
embedded into an environment with respect to which it is subsonic. This is because the CNM fragments collide supersonically. 

Now a description of how  molecular clouds are created is proposed. 
An important question is whether 
 they can be created by a single compression event from WNM
or whether  they are more  gradually created from cold dense HI clouds.
\citet{inoue2008}, \citet{inoue2009}, \citet{heitsch2009} 
and \citet{kortgen2015} have given a somewhat negative answer to the former question. 
\citet{inoue2012} have given the positive answer to the latter question. 
More detailed analyses are done by \citet{iwasaki2018}. 
Figure \ref{Fig:MagLayer} shows typical results of the compression of magnetized multiphase HI clouds by shock waves. 
The relative angle ($\theta$) between the shock wave propagation direction and the mean magnetic field is 3 degrees (upper panel),
 11 degrees (middle panel), 36 degrees (lower panel), respectively. 
The compression with a small angle results in the creation of substantial molecular gas. 
But if the relative angle is larger than a certain critical value, the propagation of shock wave only result in 
the magnetically supported HI clouds. Note that the value of this critical angle depends on the velocity of the incoming 
flow  of the magnetic field and of the time since for long enough time, 
the post-shock layer will always be at least partially molecular.
In practice, there is however a distribution of angles between the magnetic field and the incoming flow.
Quantifying this distribution would therefore be important to constraint the scenario of 
molecular cloud formation. Detailed investigations from larger scale simulations (see section~\ref{section_ls}) 
have shown that there is a clear trend for magnetic and velocity fields to be preferentially 
aligned \citep[e.g.][]{iffrig2017}, which would imply that aligned configurations are more frequent 
than expected. 

\section{The nature of MHD turbulence in the dense ISM}
Turbulence is ubiquitous in fluid dynamics and 
unsurprisingly, many observations suggest that molecular 
clouds are turbulent \citep[e.g.][]{elme2004,hf2012}. 
It is likely the case that together  with gravity, turbulence
is playing a significant role in the evolution of 
molecular clouds for example by creating 
strong density fluctuations, owing to its supersonic nature,  that may serve 
as seed for the mass reservoir of future stars. 
More generally, interstellar turbulence is an agent that imposes order in the form of 
coherent structures and correlations between the various  fields of the flows.
Turbulence  is likely responsible of many, if not almost all 
\footnote{In principle gravity and stellar feedback are two other sources whose 
signature can sometimes be clearly recognised. In many circumstances however, it is likely 
difficult to clearly separate the different contributions as gravity and feedback 
 trigger turbulent motions.}, 
of the observed motions. How magnetic field affects turbulence in molecular clouds is the main 
focus of this section.


\subsection{Turbulence in {\it ideal} MHD framework}

Before starting a description of the turbulence, an 
important issue should be stressed. {\it Ideal} MHD
implies that fluid particles are attached to their field lines, 
that is to say they can flow along the field lines but cannot 
go across them. In a turbulent fluid, given the stochastic nature of the 
motions, such a situation 
would lead to a field that would be so tangled, that quickly
motions would be prohibited.  This implies that
{\it Ideal} MHD cannot, strictly speaking, be correct for 
a turbulent fluid and that some reconnection, that is to say 
some changes of the field lines topology must be occurring.
The physical origin of this reconnection is still debated
but an appealing model has been proposed by 
\citet{lazar1999}. In this view the reconnection 
is driven by turbulence and is a multi-scale process, 
that is unrelated to the details of the  microphysical processes
\citep{lazar2015}. It is certainly the case, at least in 
numerical simulations of MHD turbulence, where the numerical diffusivity 
is often controlling the reconnection, that the MHD is far to be ideal. 
This process, in particular, induces an effective diffusion 
of the magnetic flux, that is therefore not fully frozen as
one would expect if MHD was truly ideal.

\subsubsection{Incompressible  magnetized turbulence}
\label{incompress}
For pure hydrodynamics, i.e. in the absence of magnetic field, 
the  Kolmogorov dimensional scaling relation,
 appears to provide a good description \citep{kolmo1941}.
However,  MHD flows are  more complicated and 
in spite of intensive efforts, even the energy powerspectrum 
of MHD turbulence is still debated.
The first model to predict a  powerspectrum has been done by \citet{iro1963} 
and  \citet{kraich1965} who
 infer  $v_l \propto
l^{1/4}$ and $E(k) = k^2 P_v(k) \propto k^{-3/2}$.  The power spectrum
$E(k) \propto k^{-3/2}$ is therefore shallower than the Kolmogorov
one. One of the fundamental assumptions of \citet{iro1963} 
and  \citet{kraich1965} is 
that the eddies are isotropic.
However, numerical and observational data suggest
that in MHD turbulence the energy transfer occurs mainly
 in the field perpendicular direction \citep{biskamp2003}.

An important step forward has been accomplished 
by \citet{gold1995}.
 They proposed a theory in which anisotropy of the eddies is accounted for. 
As the energy
cascade proceeds to smaller scales, turbulent eddies get more and more
elongated in the direction of  the  magnetic field. 
They assume that the Alfv\'en time-scale and the non-linear cascade
time-scale are comparable,  $k_z V_a \simeq v k_\perp$, while 
the cascade time in the direction  perpendicular to the field leads 
to $v_\perp \propto k_\perp^{-1/3}$.  The wave vector
along the z-axis is thus expressed  as  $k_z \propto k_\perp
^{2/3}$.  The energy transfer time
is therefore different from the Iroshnikov-Kraichnan estimate, and identical to
the one obtained by Kolmogorov.  One gets 
$E(k_\perp) \propto k_\perp ^{-5/3}$.  This issue has been  further studied 
  \citep[e.g.][]{cho2002,boldy2005,lee2010,beres2011,mason2012,wan2012}
and remains  still  debated. It is however clear from the numerous numerical simulations
performed that the turbulence is very anisotropic 
\citep[e.g.][]{Grappin2010}.

\subsubsection{Compressible  magnetized turbulence}
\label{compress}

\setlength{\unitlength}{1cm}
\begin{figure} 
\begin{picture} (0,10)
\put(0,5){\includegraphics[width=7cm]{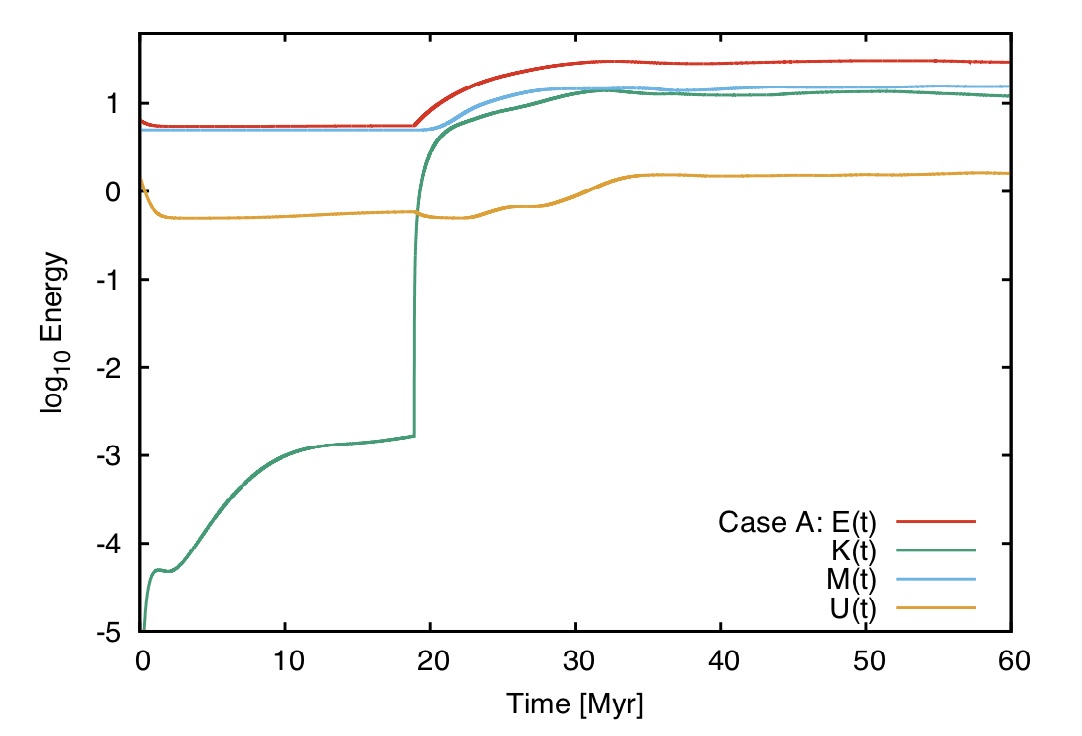}}
\put(8,5){\includegraphics[width=7cm]{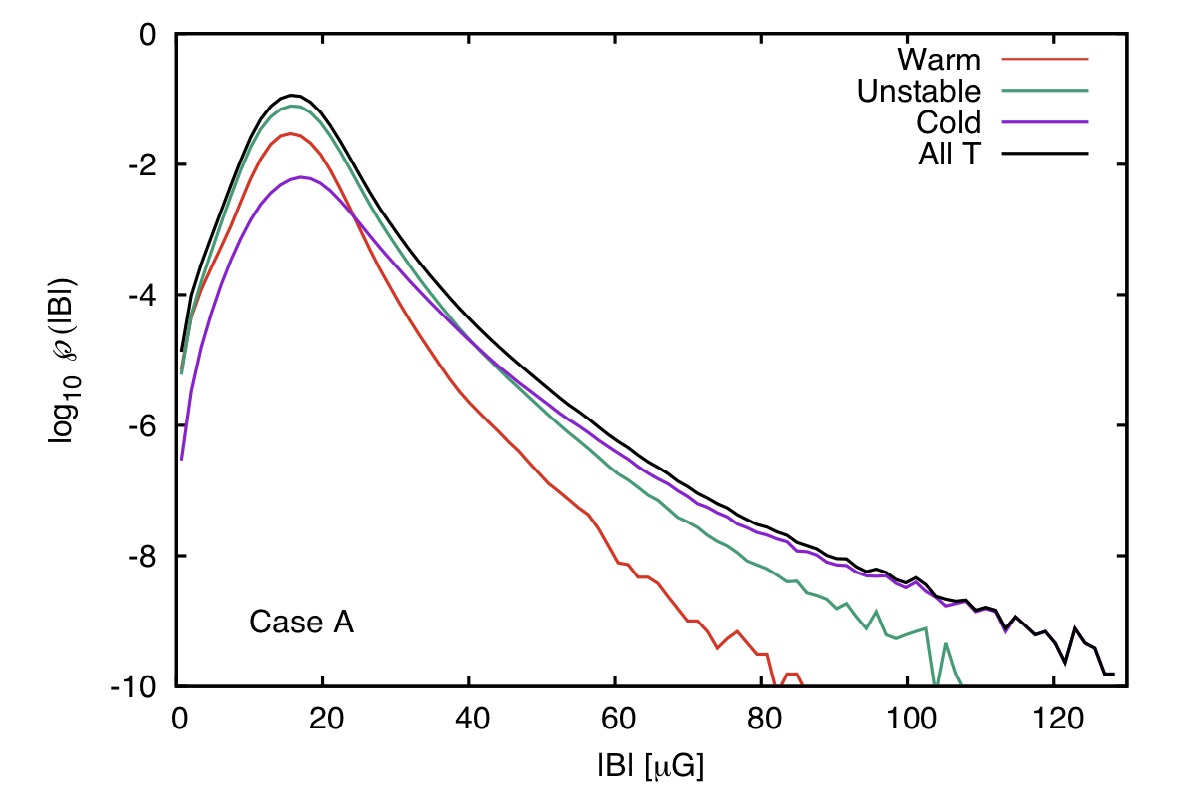}}
\put(8.,0){\includegraphics[width=7cm]{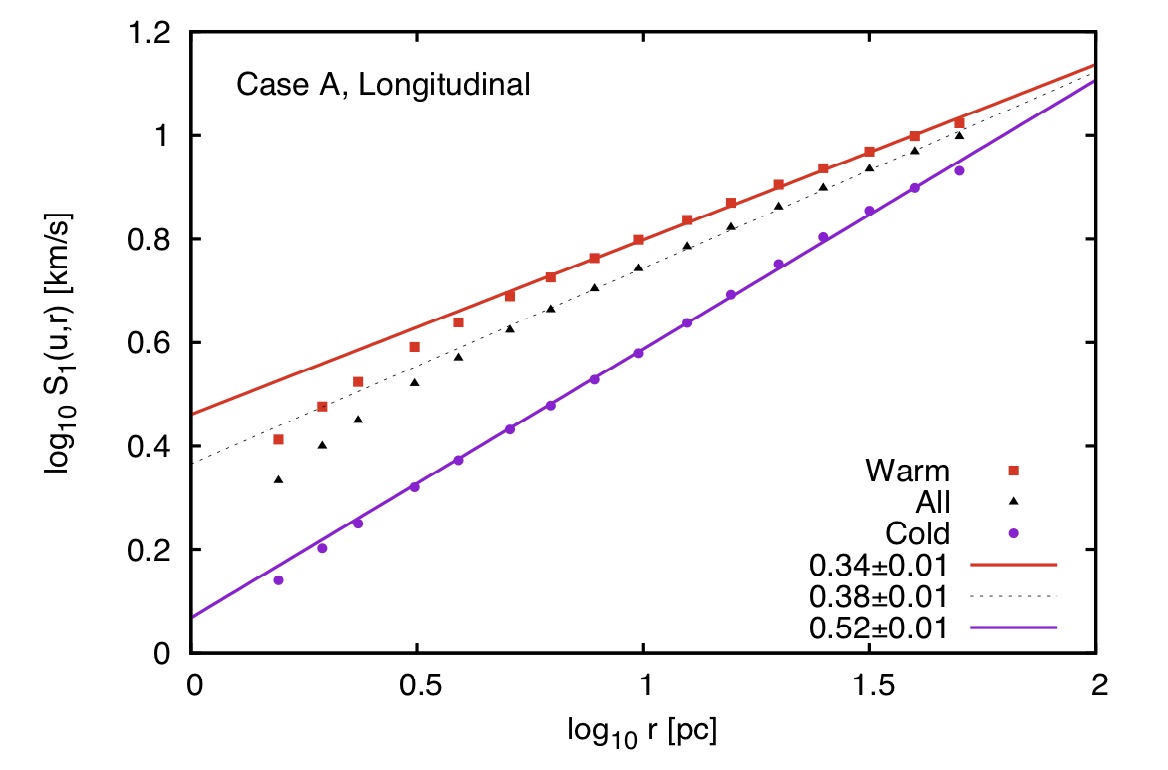}}
\put(0,0){\includegraphics[width=7cm]{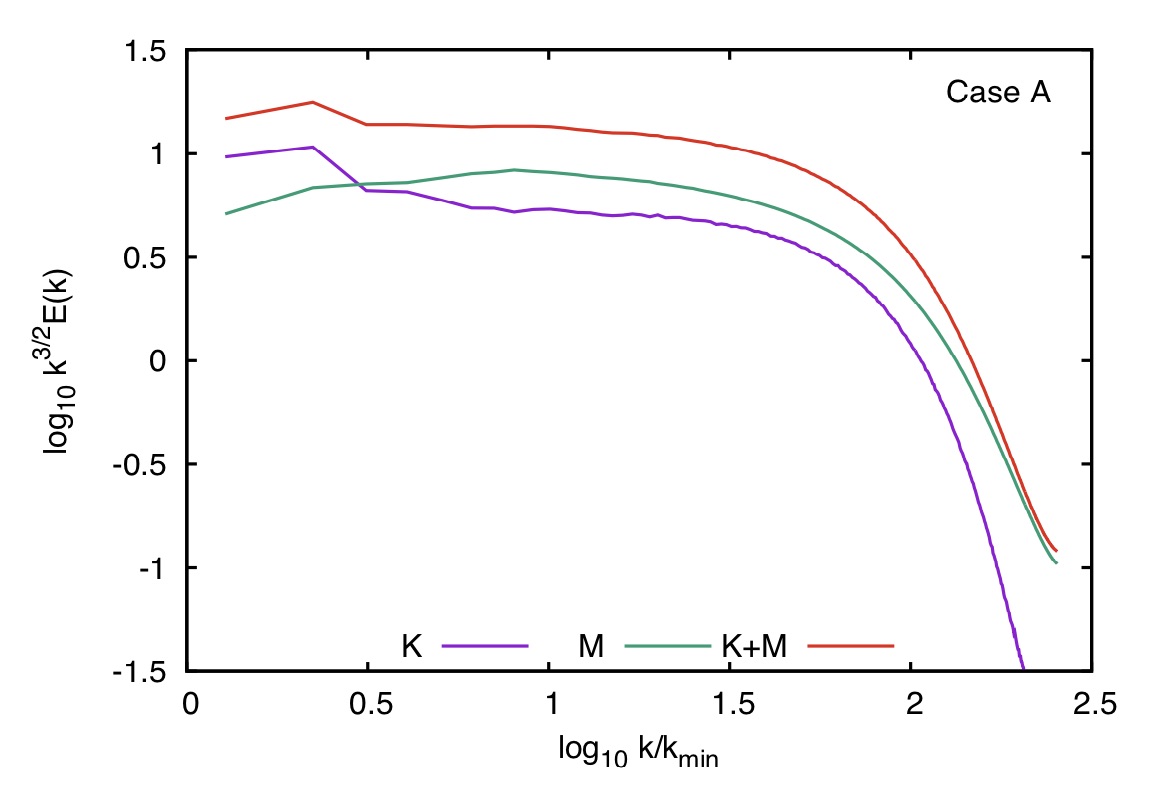}}
\end{picture}
\caption{Results of simulations of ISM magnetized 
turbulence performed by \citet{kritsuk_njp_2017}. 
Top-left panel shows the total, kinetic, magnetic and internal energies as a function of time.
Top-right panel displays the magnetic field PDF in the warm, unstable and cold phases.
Bottom-left panel portrays the powerspectra of the kinetic, magnetic and total energies while 
bottom-right shows the structure functions of the velocity in the various phases of the ISM. }
\label{krit_fig}
\end{figure}

Since molecular clouds are both magnetized and super-sonic 
(with typical Mach numbers on the order of 10), compressible 
magnetized  turbulence has received considerable attention 
during the last two decades. Because of its simplicity 
many works have been assuming an isothermal equation of state. 
More recently 2-phase medium has also been considered. 

One of the major question that has been under investigation 
when the importance of turbulence was established, was the origin of the turbulence in
molecular clouds and more precisely how the observed 
turbulence could be maintained.
Since turbulence is expected to decay in one crossing time, this 
would imply that either molecular clouds were young, either there
was a source that was continuously rejuvenating the turbulent energy, 
either the turbulence was decaying slower than expected. 
Several works have been investigating the latter assumption \citep[e.g.][]{maclow1998,ostriker2001,birnboim2018}. 
They conclude that while magnetic field introduces some delay compared 
to the hydrodynamical case, the decay still occurs too rapidly, that is to say in about one
crossing time, to explain the high level of turbulence found in 
molecular clouds.

The second major reason to study turbulence is obviously to get a statistical 
description of the fluctuations arising in molecular clouds and this has been 
addressed in several studies \citep[e.g.][]{maclow1998,padoan1999,ostriker2001,cho2003}. 
In this respect, one of the most comprehensive set of simulations  relevant for the
MHD turbulence arising in the  2-phase interstellar medium 
has been performed by \citet{kritsuk_njp_2017}. 
Five runs are presented where the mean density (2 and 5 cm$^{-3}$), magnetic field 
($\simeq$ 1, 3 and 10 $\mu$G) and 
root mean square velocity are varied. The total box size is  200 pc and a 
random forcing in the Fourier space is applied to sustain the turbulent motions. 
A cooling function relevant for the ISM is employed and it leads to the 
existence of  WNM and CNM. 
Figure~\ref{krit_fig} displays the energies as a function of time, the magnetic field PDF, 
the powerspectra of kinetic and magnetic energies as well as the longitudinal structure 
function for run A (mean density is 5 cm$^{-3}$, magnetic intensity of 10 $\mu$G and rms velocity about
16 km s$^{-1}$) of \citet{kritsuk_njp_2017}. As can be seen for this particular run the 
magnetic and kinetic energies are comparable. The PDF is  broad and magnetic intensities larger than 
100 $\mu$G are sometimes found. The energies present  power spectra with exponent compatible 
with  -3/2 although the inertial range is probably not extended enough to make this 
value well determined. Interestingly, we note that the structure function is stiffer in the CNM than in the 
WNM. Clearly this is because the former is highly supersonic while the latter is nearly transsonic. 
Let us also stress that the anisotropic nature of the MHD turbulence, which present elongated 
structures along the magnetic field as inferred in the incompressible case, 
 is still present in  the compressible  case as shown for example by \citet{vestuto2003} or \citet{Beresnyak2005}.

Due to its very non-linear nature, the description presented in most works is 
essentially numerical however
some recent theoretical progress have been accomplished for the understanding 
of how the cascade proceeds in a magnetized, compressible, self-gravitating and isothermal 
gas by 
\citet{banerkrit2017} and \citet{banerkrit2018} following the work of \citet{galtier2011}. 
In this work a complete expression of the total energy transfer is obtained as 
mixed second-order structure functions (see equation 33 of \citet{banerkrit2018}). The contributions of 
pure kinetic, magnetic, gravitational and thermodynamic terms is clearly identified
and will allow  future works to clarify their respective roles and importance.

\subsubsection{How magnetic field affects the density PDF}
The density PDF is a key quantity in the ISM,  particularly
for the star formation process. Several models 
aiming at providing explanations for the two most fundamental 
problems of star formation, namely
  the initial mass function of stars \citep{padoan1997,hc08} and  the 
star formation rate \citep{padoan2011,hf2012,fedban2015} directly depend on the 
density PDF. 

The density of
cold and weakly self-gravitating molecular gas has 
been found to present a  lognormal distribution. It is likely 
the  result of random shocks induced by the compressible turbulence and 
the multiplicative nature of the density variable leading, to a Gaussian 
distribution of $\log \rho$. A useful calculation  has been 
inferred by \citet{hopkins2013}, who derived  a 
log-Poisson distribution for the density, using 
intermittent models developed in the context of incompressible turbulence. 
The mathematical expression of the density distribution  presents  a free parameter
that controls the degree of intermittency and the 
deviation from the lognormal distribution. \citet{hopkins2013}
 compared this expression  with PDF from 
numerical simulations and obtain very good agreement.
This is particularly interesting for the high Mach number runs in which 
important deviations from the lognormal behaviour are observed.
Another important aspect regarding the cold and non-self-gravitating 
gas is the cooling or more precisely the effective equation of state, that 
is to say the pressure vs density relation. 
In most of the studies  the isothermal assumption has been made.
However powerlaws instead of lognormal have been inferred for 
polytropic flows. \citet{fedban2015} carried out  a set 
of  calculations for polytropic flows, i.e. following 
$P \propto \rho^\Gamma$ for $\Gamma=0.7$, 1 and 5/3. They inferred
modest differences between $\Gamma=0.7$ and 1 that do not 
strongly deviate from lognormal distribution. On the other hand, significant deviations were 
obtained for $\Gamma=5/3$ in particular the  low density 
part of the PDF  is better described by a powerlaw. 

The effect of the magnetic field on the density PDF has also been
studied  in the isothermal case 
\citep[e.g.][]{ostriker2001,lemaster2008}   and in 
two-phase flows \citep[e.g.][]{hetal2008,kritsuk2017}. 
It has generally been found  that
magnetic field  has a limited impact.  This  agrees
 with the conclusion that the gas which is not self-gravitating
tends to  flow along magnetic  field lines.
\citet{molina2012} carried out isothermal  simulations with
various Mach numbers. They inferred that in the transsonic and
subsonic flows, the density PDF of magnetized and pure hydrodynamical cases 
are very similar. They report however significant differences for 
supersonic flows. An analytic expression which is an 
extension of the lognormal distribution has been proposed. From their figure~1, it appears 
that the difference between hydrodynamical and magnetized runs are 
important only for the low density gas while the PDF at high densities are almost identical.

\subsection{The influence of the ion-neutral drift on MHD turbulence}

\setlength{\unitlength}{1cm}
\begin{figure*} 
\begin{picture} (0,8)
\put(0,0){\includegraphics[width=8cm]{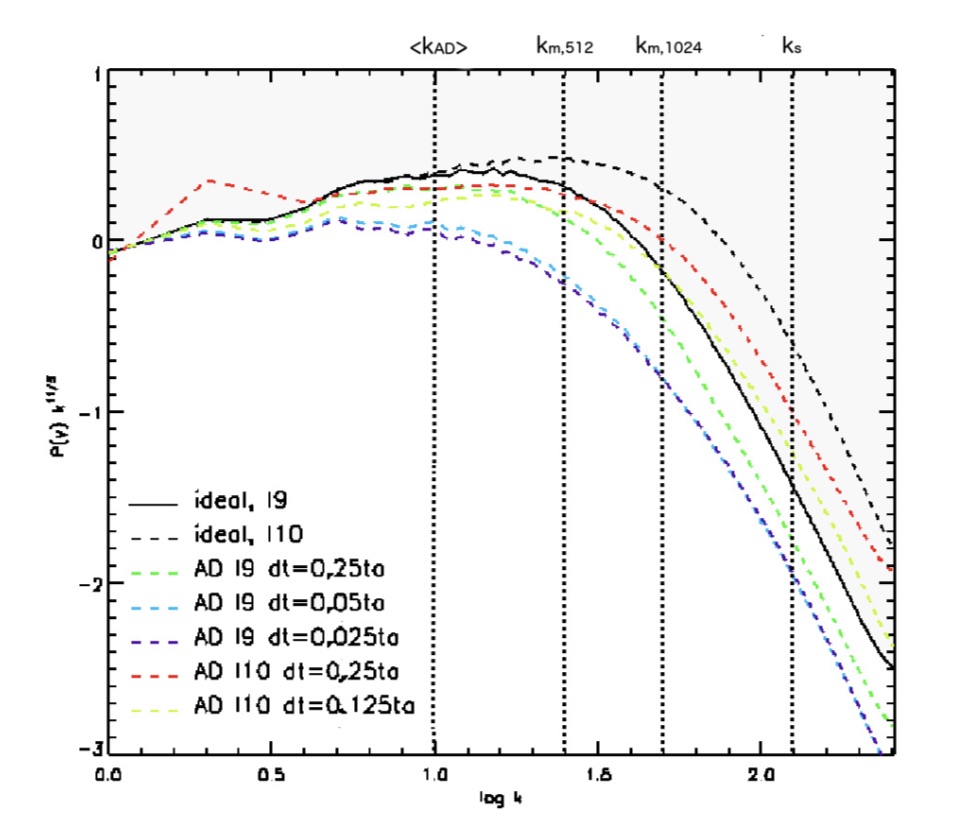}}
\put(8,0){\includegraphics[width=8cm]{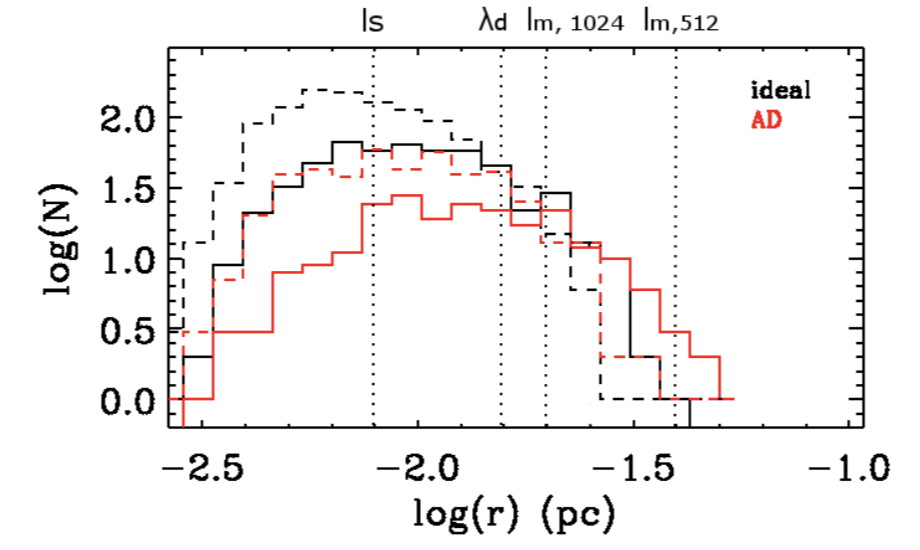}}
\end{picture}
\caption{Left panel: compensated velocity powerspectra for the decaying 
MHD simulations with ambipolar diffusion  \citep[from][]{ntormousi2016}.
The various runs include ideal MHD with 512$^3$ (l9) and 
1024$^3$ (l10) resolution and a series of runs with ambipolar diffusion with the same 
two resolutions and various values of the minimum timesteps allowed. 
The powerspectra present major deviation from the ideal MHD runs at scales
smaller than the ambipolar diffusion one. 
Right panel: distribution of filament width in MHD simulations with and without ambipolar diffusion. 
The solid lines are for a resolution of 
512$^3$ while the dashed ones correspond to 1024$^3$. 
Reproduced from \citet{ntormousi2016} with permission of A\&A.} 
\label{eva}
\end{figure*}

\setlength{\unitlength}{1cm}
\begin{figure*} 
\begin{picture} (0,15)
\put(0,10){\includegraphics[width=16cm]{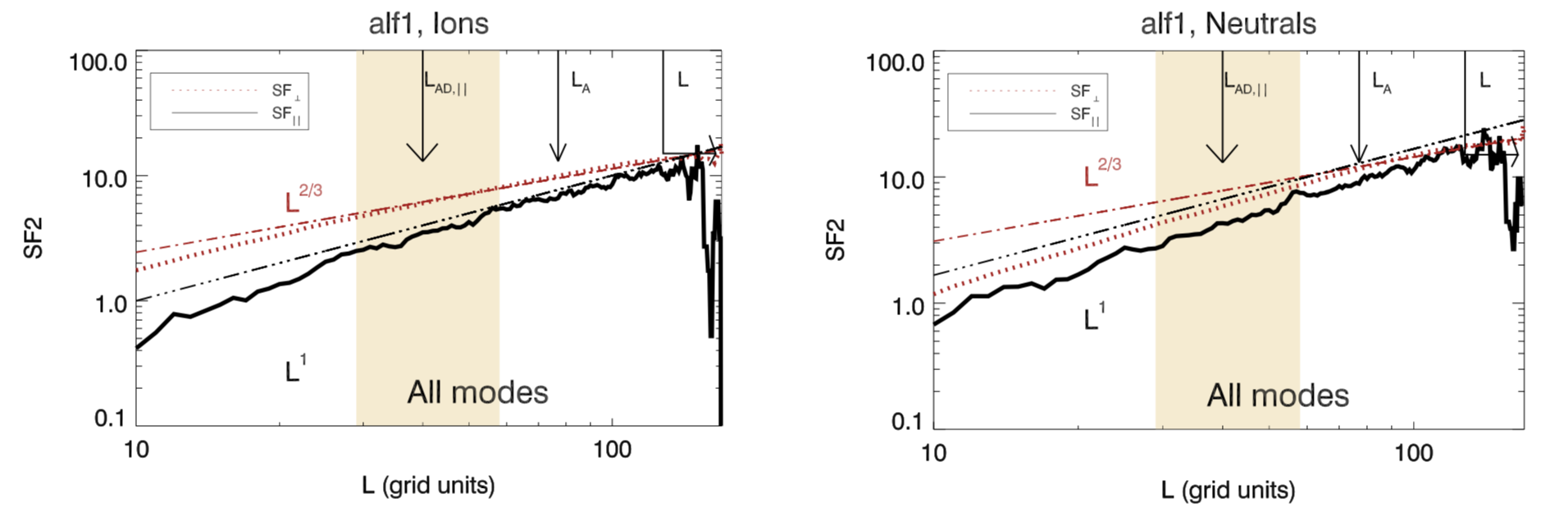}}
\put(0,5){\includegraphics[width=16cm]{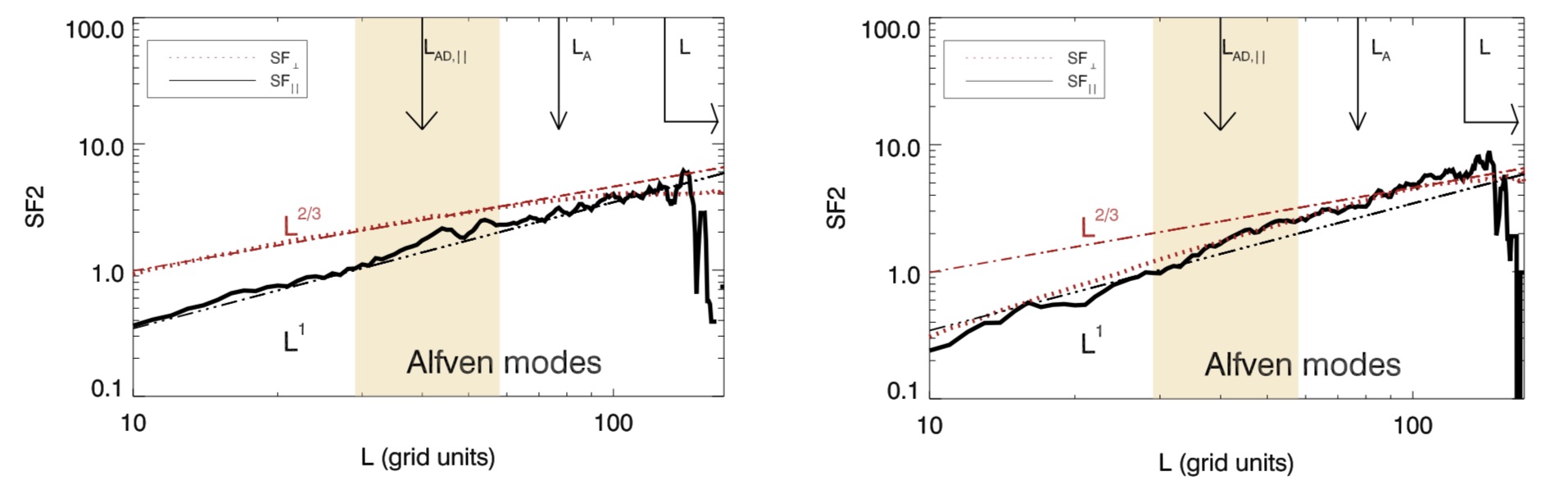}}
\put(0,0){\includegraphics[width=16cm]{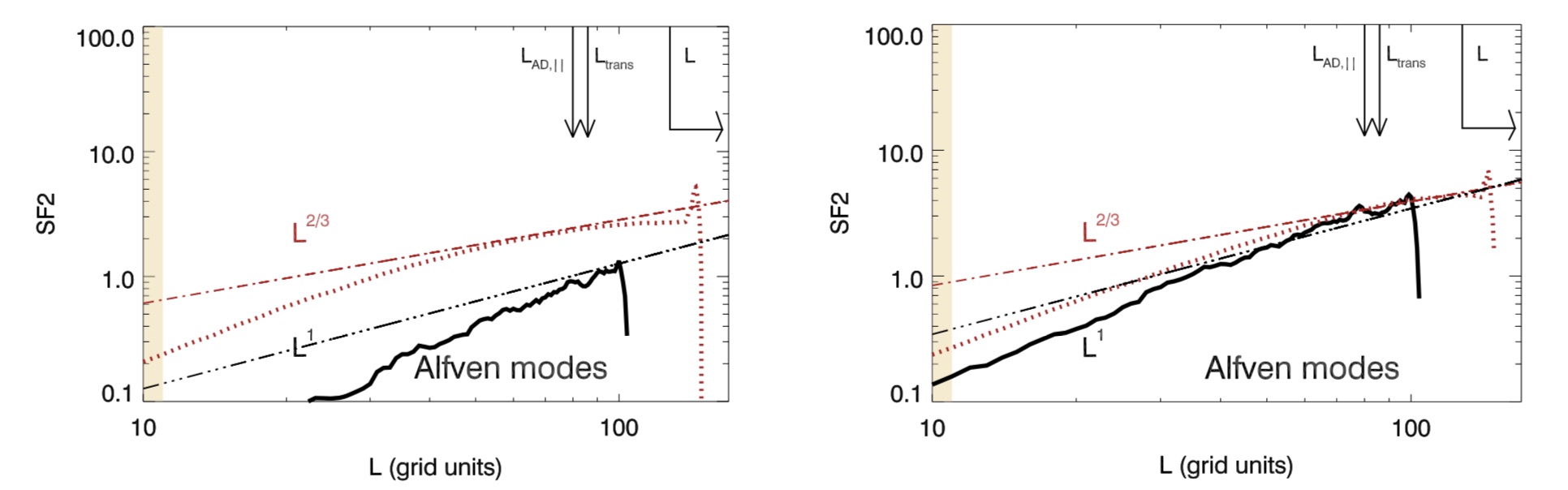}}
\end{picture}
\caption{Structure function in MHD turbulence with ion-neutral friction
from \citet{burkhart2015}. Left panels show the structure function for the ions 
while the right panels display the structure functions for the neutrals.
The first and second rows are for a supersonic and super Alfv\'enic simulation
while the third row is for a sub-Alfv\'enic one. For the second and third panel 
mode decomposition has been performed and only the Alfv\'en modes are shown. 
As can be seen from third row, they are strongly damped in the sub-Alfv\'evic case
why they roughly follow the expected scaling from ideal MHD theory in the 
super-Alfv\'enic one. 
Reproduced from \citet{burkhart2015} with permission of ApJ.}  
\label{blakesley}
\end{figure*}

As discussed in section~\ref{ambi} the ion-neutral friction is an important source of energy dissipation in 
the interstellar medium and particularly within molecular clouds. Likely enough this 
should have an impact on the development of turbulence and presumably modifies the fluctuations 
at small scales. Here we describe the various efforts that have been undertaken to investigate this 
aspect. We first describe the effects of the ion-neutral friction on MHD waves and then 
discuss the numerical simulations, which have been performed and the conclusions.

\subsubsection{How ion-neutral drift affects wave propagation}
The impact of ion-neutral friction on the propagation of  MHD waves has been 
first investigated by \citet{kulsrud1969} considering a fluid of ions and a fluid of neutrals
\citep[see e.g.][for a more recent and complete discussion]{lequeux2005}. 
Although they restrict the discussion  to Alfv\'en waves only, the dispersion relation obtained 
is of the third order making a complete discussion a little tedious.  
They found that there are 
several wavelength domains to be considered. 

In the long wavelength limit, the ions and the neutrals are well coupled because the dynamical time 
is short with respect to the ion-neutral friction time. In this limit the strong coupling 
approximation can be used and the dispersion relation is
\begin{eqnarray}
\label{disp_sc}
\omega = i {k^2 V_A^2 \over 2 \gamma_{ad} \rho_i} \pm \sqrt{k^2 V_A^2 - \left( {k^2 V_A^2 \over 2 \gamma_{ad} \rho_i} \right)^2},
\end{eqnarray}
where $k$ is the wavenumber and $V_A$ is the Alfv\'en speed of the neutrals (i.e. $V_A= B / \sqrt{4 \pi \rho}$).
The waves propagate at the Alfv\'en speed 
of the neutrals. They dissipate in a time scale that is proportional to $k ^2 \propto \lambda^{-2}$, where 
$\lambda$ is the wavelength. If $k > 2 \gamma_{ad} \rho_i / V _A$, the waves do not propagate any more.
This is because the friction is too intense.

In the short wavelength limit (which is not described by Eq.~\ref{disp_sc}), 
the waves propagate at the Alfv\'en speed of the ions, which 
for typical molecular cloud conditions, is roughly thousand times the Alfv\'en  speed 
of the neutral (because typical ionisation is on the order of 10$^{-7}$). This is because 
in this limit the wave frequency is shorter than the  ion-neutral 
friction time, thus the neutrals cannot follow the ions. The dissipation time, 
in this regime is independent of $\lambda$.

\citet{balsara1996} has been performing a complete analysis by solving for all modes and also 
by solving  for the strong coupling approximation. He concluded that the slow MHD modes
are less affected by the dissipation induced by the ion-neutral friction, particularly when 
the propagation of the waves is along the field lines. He also found that in the long wavelength  
limit, the strong coupling approximation is very accurate and can be employed. 

\subsubsection{Turbulence with ion-neutral drift}
From these analytical results, it is clear that ion-neutral friction leads to wave damping and should
therefore affect the turbulent cascade. To quantify the scale at which this may happen it is usual to 
infer the scale at which the Reynolds number, in which the viscosity is taken to be the ion-neutral 
friction, is about 1 (see section~\ref{fil_ad}). This scales is called $l_{ad}$ or $l_{diss}$, depending 
on the authors. 

One of the first simulations, that have been performed, are the ones by \citet{oishi2006}
using the strong coupling approximation. They conclude that contrary to the simple 
analytical estimate, the simulations do not reveal a clear sign of a specific 
smoothing or dissipative scale. 
Other simulations like the ones performed by \citet{limckee2008}, \citet{downes2011} 
and \citet{ntormousi2016} found that  ion-neutral friction affects the turbulent fluctuations
at a scale below the ambipolar diffusion one leading to a smoother structure. 
Left-panel of Fig.~\ref{eva} displays the velocity powerspectra 
of ideal MHD simulations at various resolution and of simulations that 
include the ion-neutral friction (for the same numerical resolutions and for various 
minimum timesteps allowed (in these calculation an explicit scheme is employed and the smallest 
timesteps is enforced by raising the ionisation if needed). Clearly the 
powerspectra with ion-neutral friction present sign of dissipation at a scale that is about 
$l_{ad}$ although numerical convergence could not be obtained. 

\citet{burkhart2015} presented three calculations with various Alfv\'enic and sonic 
Mach numbers using the heavy ions approximation. They computed 
structure functions and compare the results with the prediction made 
by \citet{gold1995}. They also performed mode decomposition as 
described by \citet{cho2003}, that is to say identifying the 
Alfv\'en, fast and slow modes. Part of their results are displayed
in Fig.~\ref{blakesley}. While the super-Alfv\'enic simulation present
structure functions compatible with the prediction of \citet{gold1995}, 
even below the ambipolar diffusion scale, $l_{ad}$, the Alfv\'en waves component of the sub-Alfv\'enic 
simulation shows clear sign of decay below $l_{ad}$. 

Clearly the nature of MHD turbulence in the presence of ion-neutral friction is 
not well understood and requires further investigation.

\section{How magnetic field correlates with the density field}
A major question to understand the role of the magnetic field in molecular cloud 
evolution is how it correlates with the other fields and in particular with the 
density. Two aspects are particularly important, first how the 
mean magnetic intensity varies with the density and second how the magnetic field 
direction correlates with structures like filaments and more generally how 
the magnetic field direction correlates with density gradients.

\subsection{The $B \, {\rm vs} \, n$ relation}
Since the pioneering work of \citet{troland1986}, it is well established 
\citep{crutcher2010} that the mean magnetic intensity is independent 
of gas density, $n$, for values up to about 300 cm$^{-3}$. At higher densities, 
that is to say at least up to 10$^{6-7}$ cm$^{-3}$, the mean magnetic intensity has been 
found to increase with $n$ broadly like an powerlaw, that is to say $B \simeq n^{\kappa}$. 
The exact value of $\kappa$ is still a matter of debate. Earlier works \citep{crutcher1999}
obtained $\kappa=1/2$ but more elaborated Bayesian analysis led to $\kappa \simeq 0.65$
\citep{crutcher2010}. 
Understanding the physical origin of this behaviour is important to unravel the star 
formation process in general. In particular, the mass to magnetic flux ratio, $M / \phi$, 
can be estimated by combining the column density of the observed component along the line
of sight and the observed magnetic intensity.
This leads to the conclusion that the atomic and diffuse molecular gas is {\it subcritical}, 
that is to say dominated by the magnetic field, while dense regions, such as dense cores, are 
generally supercritical.

Before describing the results inferred from numerical simulations, it
is worth to recall the different behaviors that can be expected.  If
the contraction occurs along the field lines, then the magnetic field
is not amplified and $B \propto n ^{\kappa}$ with $\kappa =0$. If
the motion is perpendicular to the field lines, then it is easy to
show that $n / B$ stays constant (combining the continuity and
induction equations in one dimension) and thus $\kappa=1$.  Note that in this
configuration the magnetic pressure is proportional to $n ^2$ and
therefore quickly halts any contraction. Qualitatively at least, these
two cases represent respectively a situation in which the magnetic 
field is strong and weak with respect to the kinetic motions, i.e. sub and 
super-Alfv\'enic situations. 
In the sub-Alfv\'enic case, the magnetic field guides the flow and forces the contraction 
along the field lines while in the super-Alfv\'enic 
 case, it is advected 
by the flow and the transverse component of the field is amplified.

If the contraction is
spherical, for example driven by gravity, then the mass enclosed is simply $\propto \rho R^3$, 
$R$ being the cloud radius, while
the magnetic flux is $\propto B R^2$ thus leading to $B \propto
n ^{2/3}$. It is, however, likely  that a contracting cloud does not remain
spherical, especially if the magnetic field is not negligible. In this
case, it is expected that an equilibrium along the field lines settles
leading to $c_s^2 \simeq \phi$, where $\phi$ is the gravitational
potential.  The Poisson equation leads $\phi \propto n h^2$ where
$h$ is the thickness of the cloud along the field lines. Then, as the
mass enclosed is now $\propto n R^2 h$ while the magnetic flux is
still $\propto B R^2$, we get $B \propto c_s n ^{1/2}$.  \citet{basu2000}
has compared the data provided by \citet{crutcher1999}  with this expression
and has obtained a good agreement, which improves if the velocity
dispersion $\sigma$ instead of $c_s$ is used.  Another even simpler
interpretation of this relation is energy equipartition between
magnetic and kinetic energy, $B^2/(4 \pi) \propto n \sigma^2$.

Several theoretical studies have been investigating the $B \, {\rm vs} \, n$ relation.  
In particular various simulations of 3D ideal MHD turbulence tend to show that
in realistic ISM conditions and without gravity
 \citep[e.g.][]{padoan1999,hetal2008,baner2009}, the magnetic intensity weakly depends on the
density field. A weak correlation is found with typically $\kappa
\simeq 0.1-0.2$.  This has been interpreted in the context of
the 2-phase ISM  by \citet{henne2000} as a consequence of
the magnetic tension, which tends to unbend the magnetic field lines
and to align the magnetic and the velocity fields. This eventually
facilitates the gas contraction. For polytropic flows,
the lack of  correlation is due to 
 the various types of  MHD waves having 
 different scalings of the field strength with the
density \citep{ostriker2001,passot2003,burkhart2009}.  
Indeed while for fast waves, magnetic intensity and density
are correlated, they are anti-correlated for slow waves and not
correlated for Alfv\'en waves.  Thus, in a turbulent transsonic flow, as is the multi-phase HI, the field strength is a
consequence of the complete history of wave propagation.  
Note that in supersonic and superalfv\'enic simulations, more vigorous dependence of B on the density
is inferred \citep{ostriker2001,burkhart2009}.  . 
The simulations which treat both self-gravity and
turbulence find that at high density the magnetic intensity is
$\propto n ^{0.5}$ \citep{hetal2008,baner2009}, 
which accords well with the analytical predictions deduced above.
More recently \citet{limckee2015} performed high resolution adaptive mesh simulations for a weak and a strong 
initial magnetisation and performed 
clump identification. They then investigated 
the relation between the mean magnetic field, $\bar{B}$ and the mean density, $\bar{n}$, within the clumps and 
inferred $\bar{B} \propto \bar{n} ^{0.65}$ in good agreement with the \citet{crutcher2010} result.
This may seemingly suggest that at the scale of the clumps themselves, the contraction 
is nearly isotropic.  This is in good agreement with the results reported by 
\citet{mocz2017} where simulations with a broad range of Alfv\'enic Mach number, ${\mathcal M}_A$,
 have been presented.  When ${\mathcal M}_A > 1$, the clumps follow $B \propto n^{2/3}$, while 
when ${\mathcal M}_A < 1$, $B \propto n^{1/2}$ is inferred.

\subsection{The orientation of magnetic field}
\label{orientation}
The orientation, or more generally the topology, of the magnetic field is expected to play a significant 
role in the formation of structures. 
For example as discussed above strong toroidal fields can induce
instabilities  in filaments while poloidal ones tend to stabilize them
\citep{fiege2000}.
Another example comes from the work of \citet{nagai1998}, where
 linear stability analysis of a magnetized self-gravitating
 layer has been performed \citep[see also][]{vanloo2014}. They show that the orientation of the most unstable mode tends to
 be correlated with the magnetic field direction. The result depends
 on the external pressure that determines the scale height, $z_b$ at
 which the solution is truncated. If $z_b \gg l_0$, $l_0$ being the Jeans length,
 then the fastest
 growing mode is aligned with the magnetic field, resulting in
 filaments which are perpendicular to the field direction.  The
 physical reason is that, since the width is large relative to the
 Jeans length, the layer is compressible and density fluctuations are
 easier to develop along the magnetic field.  On the other hand when
 $z_b \ll l_0$, the fastest growing mode is perpendicular to the
 magnetic field and the filaments are aligned with it.  This is
 because the layer is almost incompressible (since the scale height is
 smaller than the Jeans length), thus the instability develops through
 the bending of the layer. As perturbations whose wave vectors are
 perpendicular to the magnetic field do not bend the field lines,
 these perturbations develop more easily.

Observationally significant progresses have recently been accomplished regarding 
the magnetic field orientation. 
  The polarization observations by the Planck satellite reveal  that  in the diffuse ISM,
the elongated column density structures traced  tend to be  predominantly aligned with the magnetic field 
 within the  structures \citep{planck2016}.
This statistics for the low column density gas are comparable  to that found between 
low column density  fibres traced by $H_I$ emission and 
the magnetic field  \citep{clarks2014}.
The analysis of the Planck   data 
towards  nearby  molecular clouds  reveals that the relative orientation between 
the  structures and the magnetic field  depends on the column density, $N_{\rm H}$. 
It is  mostly parallel at $\log (N_{\rm H})  \simeq 21.7 \, \text{cm}^{-2}$ 
 and  mostly perpendicular 
at $\log (N_{\rm H}) \ge 21.7   \, \text{cm}^{-2}$ \citep{planck2016}.

\begin{figure}[ht!]
\centerline{\includegraphics[width=0.5\textwidth,angle=0,origin=c]{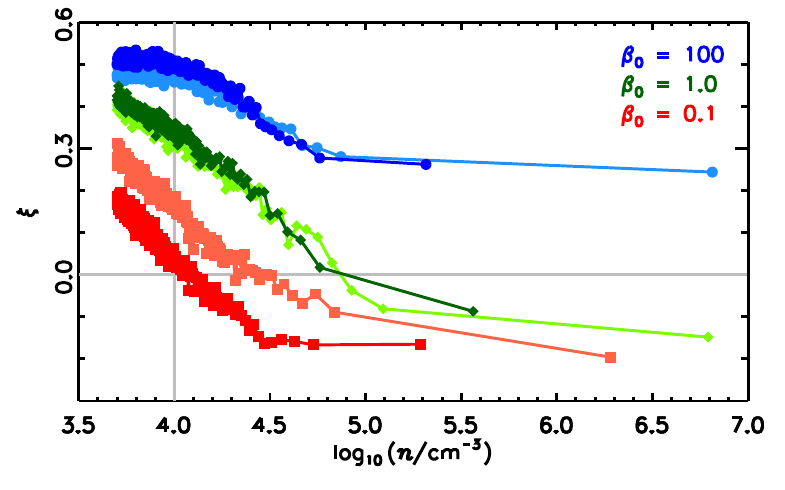}}
\vspace{-0.4cm}
\caption{Relative orientation parameter, $\xi$ vs gas density, $n\equiv\rho/\mu$ from \cite{soler2013}.
The values of $\xi$ correspond to the relative orientation between ${\bf \nabla}\rho$ and ${\bf B}$ in density bins  with $n>500$\,cm$^{-3}$.
The values $\xi > 0$ correspond to ${\bf \nabla}\rho$ mostly perpendicular to ${\bf B}$ and $\xi < 0$ correspond to ${\bf \nabla}\rho$ mostly parallel to ${\bf B}$. The grey horizontal line is $\xi=0$.
The darker colours represent the early snapshots and the lighter colours, the later snapshots.
The grey vertical line, drawn for reference, corresponds to $n=10^4$\,cm$^{-3}$.
Reproduced from \citet{soler2013} with permission of A\&A.} 
\label{fig:zeta-lognh}
\end{figure}

A detailed analysis of the angle, $\phi$ between the magnetic field and 
the density gradient,   
\begin{equation}\label{eq:relorientation0}
\cos\phi = \frac{{\bf \nabla}\rho \times {\bf B}}{|{\bf \nabla}\rho|\,|{\bf B}|},
\end{equation}
in 
 numerical simulations has been carried out by \citet{soler2013} using the simulations
 presented in \citet{dib2010}. In these numerical experiments the gas is isothermal and 
the turbulence, seeded initially with an initial Mach number of about 10, is decaying. Through shocks and self-gravity 
dense clumps and filaments  quickly form. Three values of magnetisation, characterised by 
the initial $\beta$, equals to the thermal over magnetic pressure have been explored
namely 100 (weakly magnetized), 1 and 0.1 (strongly magnetized). 
A value of $\xi > 0$ means that the dominant configuration is $\cos \phi \simeq 0$, that 
is to say the magnetic field and the density gradient tend to be perpendicular, which in turns 
implies that the magnetic field and density isocontour tend to be parallel.
To quantify the alignment, in each density bin,  the difference between the numbers of cells having 
respectively 
$| \cos \phi| < 0.25$  and $| \cos \phi| > 0.75$ \citep[see][]{soler2013} has been computed.
Figure~\ref{fig:zeta-lognh} shows the dependence of $\xi$ with the gas density for the three runs 
at two timesteps.
For the low magnetisation ($\beta=100$), $\xi$ remains positive for all density bins, 
with a clear trend for $\cos \phi$ to increase at large densities (i.e. $\phi$
 goes to smaller values). For the more magnetized case, $\xi$ becomes negative 
 at high densities and the density value at which this happens drops with $\beta$. 
This in particular shows that at low densities the magnetic field tends to be aligned with the 
filaments while at high densities it is more perpendicular to them. While the physical origin 
of this last trend is simply that the gas is channeled by the magnetic field, when it is strong 
enough, along the field lines, the mechanism by which the alignment occurs at low density 
is less obvious. To better understand it, \citet{soler2017} have obtained 
an exact equation for the evolution of $\phi$. It is simply obtained by combining the Faraday 
and continuity equations, 
\begin{equation}\label{eq:costheta3}
\frac{d(\cos\phi)}{dt} = \frac{\partial_{i}(\partial_{j}v_{j})}{(R_{k}R_{k})^{1/2}}[-b_{i}+r_{i}\cos\phi]+\partial_{i}v_{j}[r_{i}r_{j}-b_{i}b_{j}]\cos\phi,
\end{equation}
where
\begin{equation}\label{eq:defri}
r_{i} \equiv \frac{R_{i}}{(R_{k}R_{k})^{1/2}}, \; R_i = \partial _i \rho, 
\end{equation}
\begin{equation}\label{eq:defbi}
b_{i} \equiv \frac{B_{i}}{(B_{k}B_{k})^{1/2}},
\end{equation}
By numerically estimating the different terms in the right-hand side of Eq.~(\ref{eq:costheta3}), \citet{soler2017} showed 
that the mean value of the first term, which entails second spatial derivatives, quickly goes  to zero. They therefore
concluded that the second term is mainly responsible of the evolution of $\cos \phi$. As can be seen this 
second term vanishes either if  $B$ and $\nabla \rho$ are orthogonal, in which case $\cos \theta=0$, or if they are 
parallel, in which case $r_i r_j  - b_i b_j =0$. This could suggest that $\cos \phi$ has two attractors, 0 and $\pm$1
although this obviously depend on the sign of the velocity derivatives. Numerical estimates using MHD simulations, 
show that indeed, on average, the mean value of $\cos \phi$ follows the sign of the velocity terms. 

Therefore the aligned configurations ($\phi=0$ or $\pi$) and perpendicular ones ($\phi=\pi/2$) are favored. They are simple 
consequences of the fluid equations (more precisely continuity and Faraday equations).

\setlength{\unitlength}{1cm}
\begin{figure*} 
\begin{picture} (0,9)
\put(0,0){\includegraphics[width=9cm]{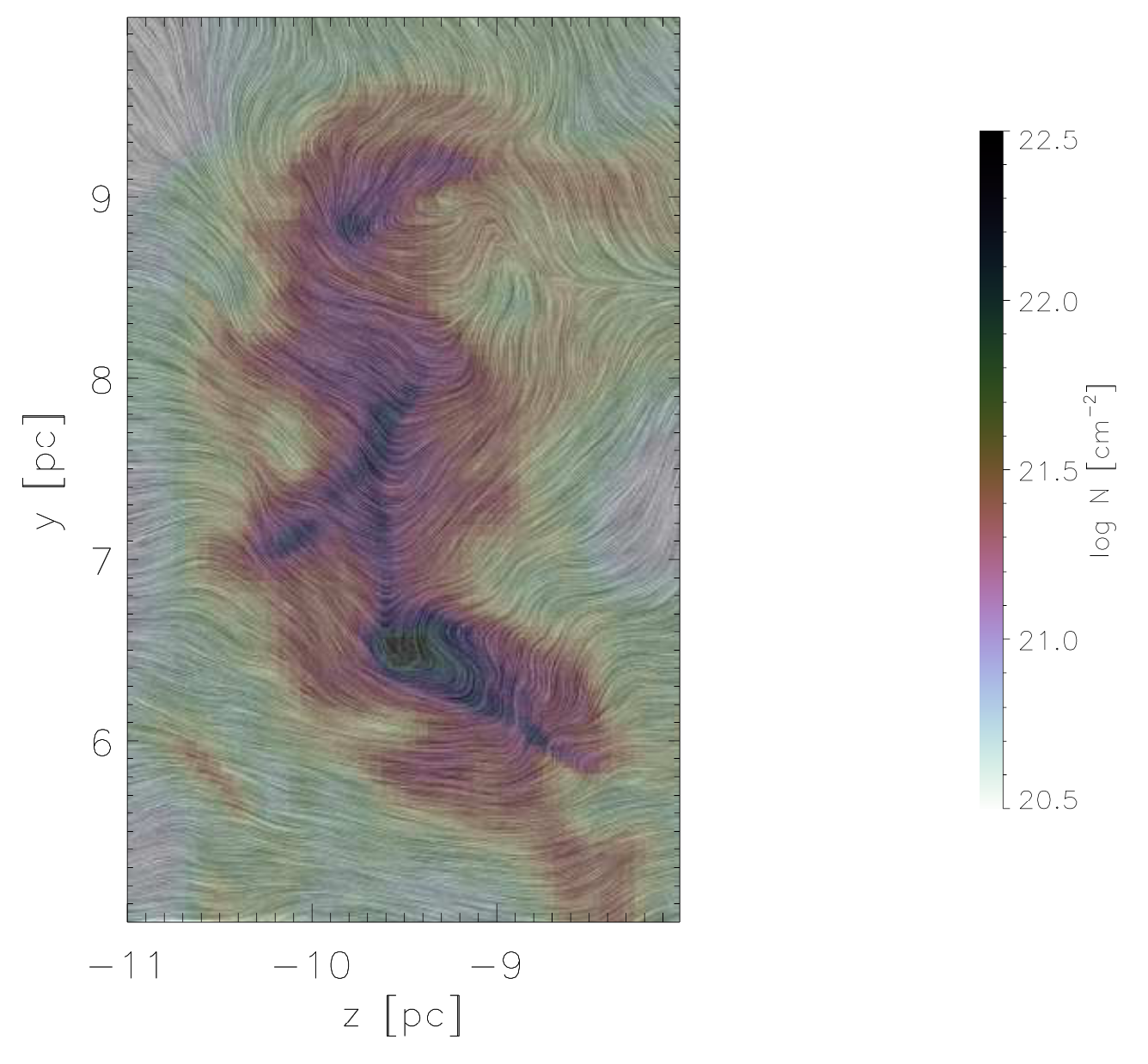}}
\end{picture}
\caption{Structure of the magnetic field within a self-gravitating filament 
\citep{gomez2018}. The structure of the mass weighted magnetic field integrated 
along the line of sight is shown on top of the column density. The magnetic field, 
that is primarily perpendicular to the direction of the filament, is then further stretched 
by the collapsing motions along the filament, resulting in a ``U''-shaped magnetic field line 
geometry. }
\label{gomez}
\end{figure*}

Recently \citet{gomez2018} have investigated the detailed structure of the magnetic field inside 
a self-gravitating filament, which forms in a turbulent environment. Similarly to 
other studies, they found that the magnetic field is primarily perpendicular to the 
supercritical filaments. However, they note that due to the gravitational infall along the 
filament, the field lines are further bent resulting in a ``U''-shaped  magnetic
field line geometry.  An equilibrium eventually  settles due to the  diffusion processes, 
that  equilibrate with the  transport by the infall motions. Figure~\ref{gomez} displays 
the magnetic field structure on top of the column density within the filament.

\section{Filaments}
While the density PDF provides very important information on the ISM, it should be kept in mind that they miss an essential piece of information, i.e. the spatial correlations or the shape of the interstellar clouds. 
While it has been recognized that the ISM is remarkably filamentary for many years, recent studies carried out by Herschel led to quantitative statistical estimates of their properties \citep{andre2014}. 

The first question that has to be addressed is what is the origin of this ubiquitous filamentary structure? 
Second, Herschel studies have also revealed that the filaments have a possible characteristic width of about 0.1 pc, which is surprising and needs to be explained although it is worth stressing that this result has for now been obtained only in 
nearby molecular clouds. 
Finally, it seems that most star forming cores sit inside self-gravitating filaments \citep{poly2013,kony2015}, seemingly suggesting that filaments may be one important step of the star formation process.

\setlength{\unitlength}{1cm}
\begin{figure*} 
\begin{picture} (0,9)
 \put(0,0){\includegraphics[width=9cm]{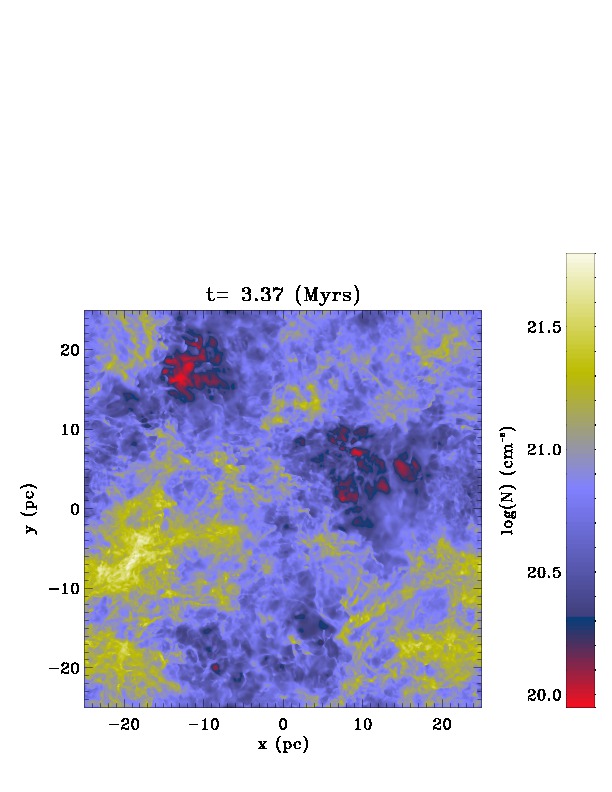}}
 \put(9,0){\includegraphics[width=9cm]{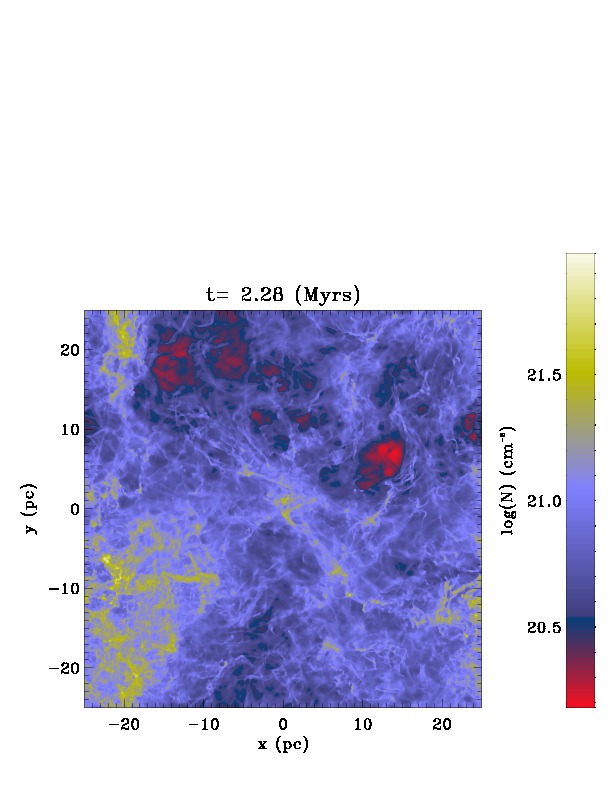}}
\end{picture}
\caption{Column density for one
 snapshot of a  decaying
turbulence experiment. Left hydrodynamical run, right  MHD run
(from \citet{h2013})..
Initially the field is uniform and has an intensity of 5$\mu$G.
The magnetised run presents a  more filamentary structure 
that the hydrodynamical run as can seen through visual inspection and confirmed by detailed analysis
(see Fig.~\ref{triaxis}).
Reproduced from \citet{h2013} with permission of A\&A.} 
\label{decay_mhd}
\end{figure*}

\setlength{\unitlength}{1cm}
\begin{figure} 
\begin{picture} (0,7)
 \put(0.,0){\includegraphics[width=8cm]{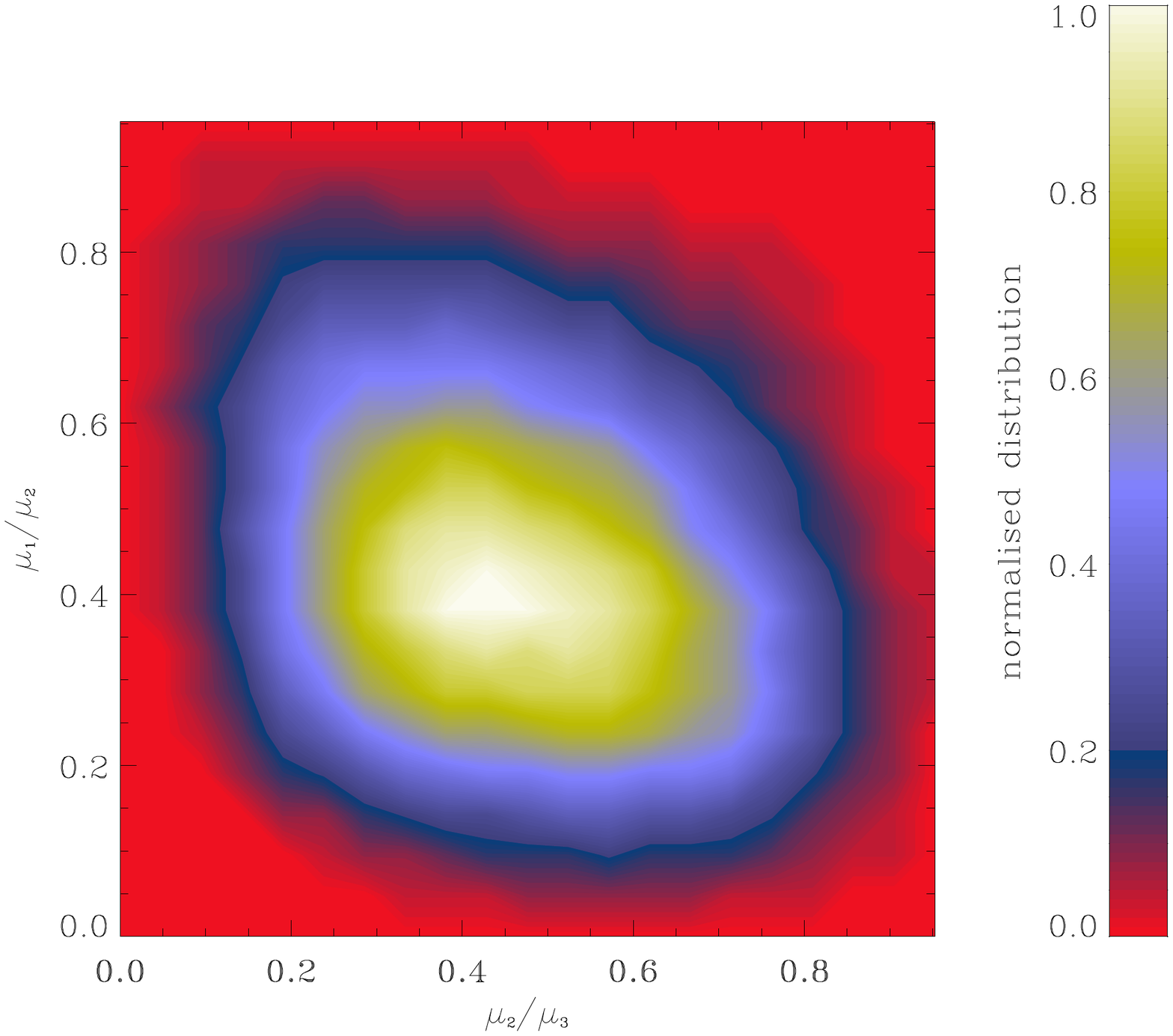}}
 \put(8.,0){\includegraphics[width=8cm]{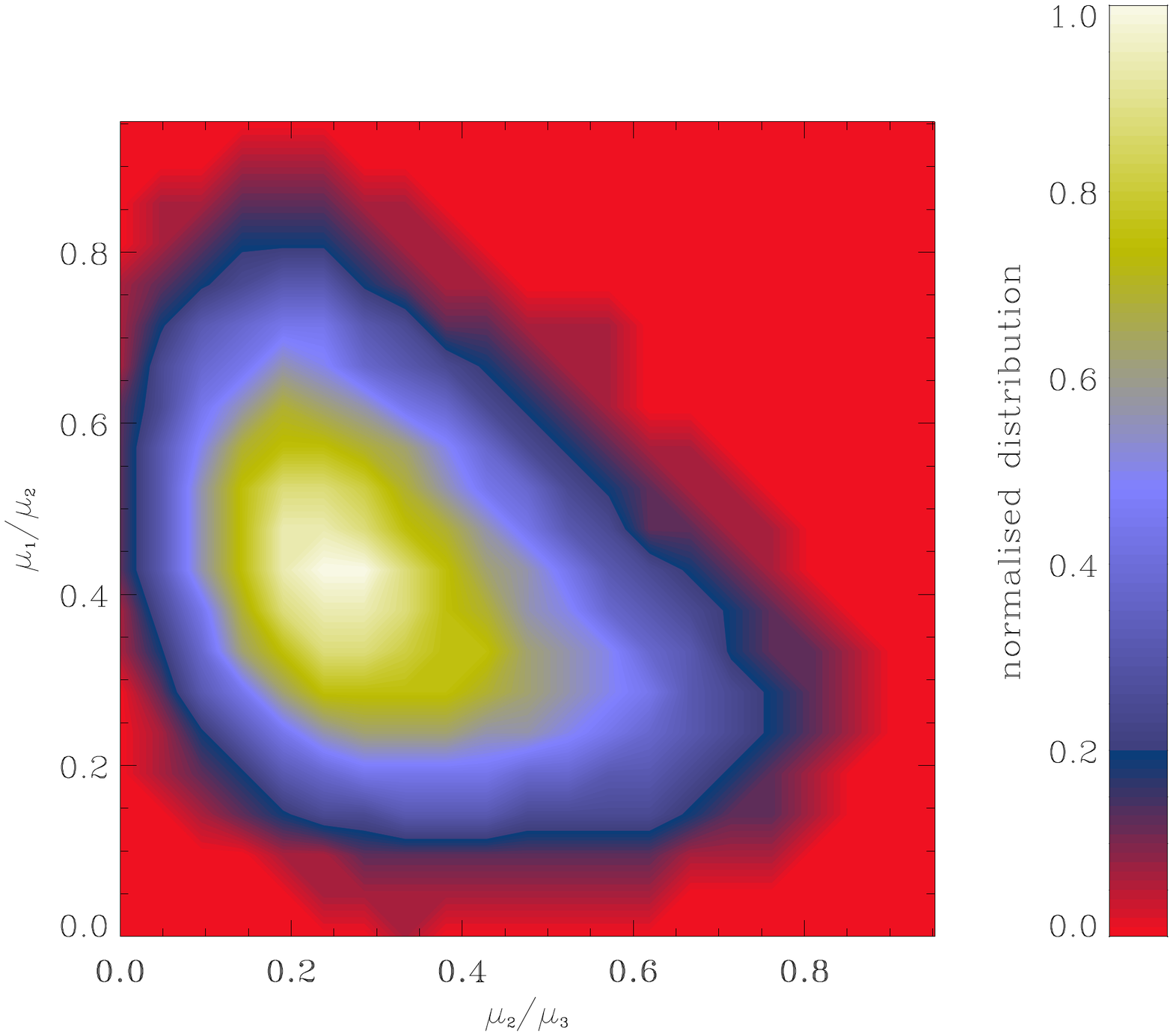}}
\end{picture}
\caption{Normalized bidimensional histogram displaying 
$\mu_1/\mu_2$ as a function of $\mu_2/\mu_3$ where $\mu_1$, 
$\mu_2$ and $\mu_3$ are  the 
inertia matrix eigenvalues, $\mu_1$ being the smallest (from \citet{h2013}).
 Left panel: hydrodynamical simulation.
 Right panel: MHD simulation. 
Clearly the MHD run present structures that on average tends to be more elongated 
(i.e. have a smaller $\mu_2 / \mu _3$) than the hydrodynamical ones.
Reproduced from \citet{h2013} with permission of A\&A.} 
\label{triaxis}
\end{figure}

\setlength{\unitlength}{1cm}
\begin{figure*} 
\begin{picture} (0,9)
 \put(0,0){\includegraphics[width=9cm]{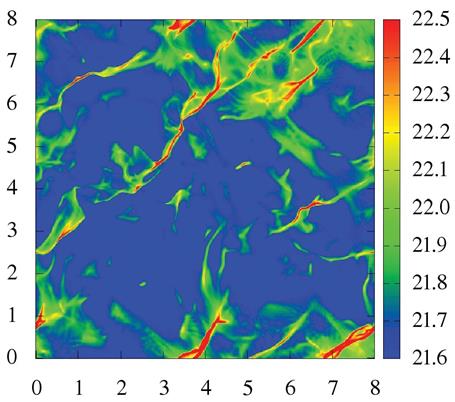}}
\end{picture}
\caption{Formation of supercritical filaments and striations \citep{inutsuka2015}
in a shocked layer (seen face on). In this calculation the filaments are self-gravitating
(and therefore named supercritical) and the striations are mainly perpendicular to the filaments.
Reproduced from \citet{inutsuka2015} with permission of A\&A.} 
\label{im9}
\end{figure*}

\subsection{Formation of filaments} 
It is well known that gravity amplifies anisotropies and tends to promote the formation 
of filaments. In the context of molecular clouds this is particularly 
evident in studies like the ones performed by \citet{nagai1998}, \citet{smith2014}, \citet{gomez2014}, \citet{fed2015}, 
\citet{gong2015}, \citet{chen2015} and \citet{camacho2016}. 
This is simply because the gravitational force being the gradient of a scalar, it is stronger along the shortest axis of a clump. 
However, gravity can not explain all the observed filaments because many filaments are not self-gravitating. Indeed, the atomic gas (HI) is itself rather filamentary \citep{miville2003,mcclure2006,clarks2014}, but is far from being self-gravitating. 
It seems therefore that other processes could lead to filament formation. 
To investigate this issue \citet{h2013} performed MHD and hydrodynamical turbulent simulations of the ISM and computed the clump aspect ratio. 
These simulations include interstellar cooling and therefore present a 2-phase structure as described above. 
Gravity is not included. 
They have an initial velocity dispersion which corresponds to a Mach number of 10 and then decay.  
\citet{h2013} found that magnetic field makes the clumps  more filamentary as  seen from 
Figs.~\ref{decay_mhd} that portray the column density in an hydrodynamical and an MHD run.
This is indeed quantified by Fig.~\ref{triaxis}, which
 shows the aspect ratio of the clumps that have been extracted from the simulations 
using a friend of friend algorithm. It reveals that on average the clumps in the MHD case have a smaller $\mu_2 / \mu_3$, 
where $\mu_2$ and $\mu_3$ are the inertia matrix eigenvalues of the clumps. 
\citet{h2013} also found  that the filament axis tends to be aligned with the strain, i.e. the direction along which the fluid particles are stretched by the velocity field. This suggests that indeed turbulence, and even more likely, MHD turbulence naturally produces elongated structures. 
This is in good agreement with the anisotropic nature of MHD turbulence which, as discussed in Sections~\ref{incompress} 
and~\ref{compress}  produces structures elongated along the magnetic field. 

\citet{InoueInutsuka2016} demonstrated that, in shock compressed layers of typical magnetized ISM, filamentary cold HI clouds are naturally created by thermal instability, and 
they also showed that stretched HI filaments that align with the local magnetic fields are due to the turbulent shear strain induced at the shock front.
Prominent filaments are also found in magnetized shock-compressed dense layer \citep{inutsuka2015,ntormousi2017}, as illustrated in Fig.~\ref{im9}, while again the unmagnetized runs produce much less elongated structures \citep{ntormousi2017}, seemingly suggesting that the effect is generic and not sensitive to particular configurations. 
In the case of shock-compressed molecular layers denser than HI clouds, the massive filamentary clouds are perpendicular to the mean magnetic field lines. 
The mechanism to create such a prominent feature can be interpreted as the generic interaction of a shock wave and a magnetized medium with significant density inhomogeneity in pre-shock state \citep{InoueFukui2013,Inoue+2018}. 
This mechanism does not require  self-gravity but the latter enhances the accretion of gas along the magnetic field lines onto the massive filament.
Note that in some calculations \citep[e.g.][]{inutsuka2015}, substructures connected, and often perpendicular, to the main filaments are also observed.
These structures are reminiscent of the {\it striations} that have been reported in molecular clouds \citep[e.g.][]{heyer2016}, where they appear as highly elongated along the magnetic field.
\citet{tritsis2016} have performed several numerical calculations to investigate their origin and concluded that they are most likely a consequence of non-linear MHD waves due to inhomogeneous density fields. Similar conclusion has been reached by 
\citet{chen2014} who presented a series of magnetized simulations and identified a network of small filaments 
aligned with the magnetic field in the simulations with the lowest $\beta$ and estimate that this later must be $<0.2$ 
to get prominent striations. 

The role of the magnetic field in the formation of filaments is likely important because it makes the flow more coherent therefore allowing the existing filaments to survive longer. 
In a related way, the flows tend also to be more organized when they are magnetized. 
For example  several studies have concluded that velocity and magnetic field are preferentially aligned \citep[see for example][]{matthaeus2008,baner2009,iffrig2017}. 
This is also consistent with the recent finding that the magnetic field direction and the density gradients are clearly correlated \citep{soler2013,koch2013,koch2014,planck2016,soler2017} as discussed in Sect.~\ref{orientation}.

Let us reiterate that there is not necessarily a unique mechanism that leads to the formation of filaments. 
In particular it is clear that both magnetic field and self-gravity tend to produce highly elongated structures. 
While it seems difficult to invoke the latter in the formation of diffuse filaments, it very likely plays a determinant role in the formation of the most massive ones. 
This is particularly obvious in series of simulations presented by \citet{federrath2016} where filaments can form under the influence of gravity and MHD turbulence only.

\subsection{A characteristic width ?}
Perhaps the most intriguing and recent aspect of filaments is the possible existence of a  characteristic width, 
of about 0.1 pc, and 
even more surprising is the fact that this remains true for filaments of column 
densities spanning almost 3 orders of magnitude \citep{arzou2011,andre2014,koch2015}.  
Indeed both gravity and turbulence tend to be scale free processes and  usually 
produce powerlaws. For example the Jeans length varies by more than one order
of magnitude in the above mentioned filament sample. 
This analysis on the width distribution in Herschel observation has triggered many subsequent papers on this issue. 
While \citet{Juvela+2012}, \citet{AlvesDeOliveira+2014} and \citet{koch2015} essentially confirmed the earlier findings, 
\citet{gina2017} pointed out the tension between the characteristic width and the spatial power spectra of the data 
that show no characteristic scale.
This tension could be removed by the fact that the masses in the filaments with a characteristic width corresponds 
to small fraction of the total mass in the whole molecular clouds and hence they provide small contribution in the 
spatial power spectra in observational emission maps \citep{Roy+2019}. 
\citet{Hacar+2018} showed the velocity coherent filamentary structures (so-called fibers) have a median widths that is a 
factor of three smaller than 0.1pc in the integral shape filament in Orion using ALMA observation of molecular emission 
from ${\rm N}_2{\rm H}^+$ (1-0). 
However, \citet{Clarke+2018} cautioned about line-of-sight confusion in the analysis of velocity coherent structure 
according to their synthetic observations of simulated filaments. 
Note also that high-resolution ALMA observation has not yet been 
reported for dust continuum emission that has a dynamic 
range in emission much larger than in molecular line observation. 
These observations lead to  the  question of why we tend to observe a characteristic width of molecular filaments, at least, 
apparently with the spatial resolution typical in Herschel observation. 
Part of the answer   may be the  finite resolution  has  
recently   claimed by \citet{gina2017}. 
This may  account for some of the observed filaments in particular the low column density ones that are not 
as prominent as the very dense ones. 

Various explanations \citep{andre2014} have been put forward to account for this fact, 
three of them are  described below.

\subsubsection{Jeans length and self-gravitational equilibrium}
It would appear logical that the width of supercritical 
filaments is directly related to the mean Jeans length within the 
filament. However, as stressed by \citep{arzou2011}, the 
Jeans length drops with density which is at odds from the nearly 
constant width that is inferred from observations.
However Fischera \& Martin (2012) argued that the characteristic size 
of the filaments is simply the result of mechanical 
equilibrium in the radial direction. Assuming that the filaments 
are pressure bounded, they  find that the equilibrium of the isothermal 
gas, between thermal pressure and gravity leads to a diameter 
of about 0.1 pc with a  weak dependence on the column density. 
While this explanation could be valid for nearly critical filaments and is indeed observed in numerical simulations \citep{smith2014}, it cannot account for very supercritical filaments as thermal support is unable to resist gravity.
The effects of magnetic field are also studied by \citet{tomisaka2014} and \citet{auddy2016} 
where a bidimensional equilibrium is considered, with the filaments being rather ribbons due to the anisotropic Lorentz force. 
Again the finite width of the massive filament cannot be explained even with magnetic field, unless the strength of the field is exceptionally large.

\subsubsection{The sonic length argument}
 If filaments are produced in shocks, then their
density, $\rho_f$, should be linked to the background density,
$\rho_0$ by the Rankine-Hugoniot relation: $\rho_f = \rho_0 {  M}^2$, 
where ${ M}$ is the Mach number, ${ M}= v / c_s$.
The velocity on the other hand is linked to the scale as 
$v(L) \simeq v_0 (L/1 {\rm pc})^{\eta}$ which is simply the Larson relation
discussed above and $v_0 \simeq 0.8 \; {\rm km \; s^{-1}}$ while 
$\eta \simeq 0.4-0.5$.  As the size of the shocked layer is simply given by
$L_f = L \rho_0 / \rho_f$, we get $L_f = (c_s/v_0)^2 \times L (L/1 {\rm pc})^{-2 \eta}$.  
In particular assuming that $\eta=0.5$,
       the shocked layer becomes independent of the fluctuation size
       and with $c_s \simeq 0.2$ km s$^{-1}$, we get $L_f \simeq 0.07$
       pc which is close to the thickness inferred by \citet{arzou2011}.
This explanation has been generalised to the magnetic case by \citep{fed2015}, who argue that it reproduces the simulations well. 
Note, however, that this argument only explains the thickness of the sheet-like structure 
that is geometrically different from the filament. 
In addition, this explanation neglects the effect of self-gravity, and hence, cannot explain why the massive filaments are supported against excessive self-gravitational forces that 
are expected to trigger the radial collapse of the filaments. 

The model considered by  \citet{auddy2016}  is more elaborated as they considered a 2D equilibrium with magnetic field lines perpendicular
to its surface. Along the field lines the structure, which is 
described as a ribbon, is really narrow and typically below  0.1 pc, while perpendicularly it is confined by the ram pressure and 
its length, close to 0.1 pc, is essentially the sonic length.

\subsubsection{The ion-neutral friction}
\label{fil_ad}
A third class of explanations has invoked the ion-neutral friction that 
provides a source of dissipation, the  ion-neutral drift presents
a  characteristic time namely $\rho_i \gamma$.
From Eq.~(\ref{induc})
a magnetic Reynolds number \citep[e.g.][]{mckee2010,h2013} can be inferred
\begin{equation}
R_{e,m}  = { V(l) l \over \nu },
\end{equation}
where $\nu = B^2/ (4 \pi \gamma_{ad} \rho \rho_i)$.
Assuming that the energy 
flux, $\epsilon = \rho V(l)^3 / l$, is constant through the scales, one gets
\begin{equation}
R_{e,m}  = { \epsilon^{1/3} \rho^{-1/3} l^{4/3} \over \nu }.
\end{equation}
Estimating $\epsilon$ at the integral scale, $L_0$, we obtain 
\begin{equation}
R_{e,m}  = \left( { \rho _0 \over \rho } \right)^{1/3} {V_0 \over L_0^{1/3}} {  4 \pi \gamma_{ad} \rho \rho_i  \over B^2 } l^{4/3}.
\end{equation}
The smallest scale that can be reached in a turbulent  cascade is typically obtained when the Reynolds number is 
equal to about 1. This leads for $l_{diss}$, the dissipation length, the following expression:
\begin{equation}
l_{diss}  = \left( {  L_0^{1/3}  \over \rho _0^{1/3} V_0 } \right)^{3/4}  \left( { B^2 \over  4 \pi \gamma_{ad} \rho^{2/3} \rho_i  } 
\right)^{3/4}.
\label{ldiss}
\end{equation}
Typical values for the ISM are
 $V_0=2.5$ km s$^{-1}$, $\rho_0=100$ cm$^{-3}$ and $L_0=10$ pc
The magnetic intensity is  about 5 $\mu$G in the diffuse gas and 
10-20$\mu$G in the 
molecular gas for densities of a few 10$^3$ cm$^{-3}$. 
In the molecular gas the ionization is  about $10^{-6}-10^{-7}$ 
\citep{Lepetit2006,bergin2007}
and the ion density $\rho_i$ is given by  $C \sqrt{\rho}$, where
 $C=3 \times 10^{-16}$ cm$^{-3/2}$ g$^{1/2}$.
 For a density of $10^3$ cm$^{-3}$, a magnetic intensity
 of $20 \mu$G, this leads to
$l_{diss} \simeq 0.2$ pc.
Obviously eq.~(\ref{ldiss}) depends on physical parameters such as $V_0$ and $B$ and therefore should present 
variations. It is worth realising that the first term of the right-hand side is the energy flux to the power $1/4$.
The energy flux, at least in Kolmogorov theory, is expected to be constant through scales. The second term may 
also present weak variations since $\rho _i \propto \rho^{1/2}$, it is proportional to $(B^2 \rho^{-7/6})^{3/4}$
while observations reveal that $B \propto \rho^{1/2}$ is not a bad approximation \citep{crutcher1999}. 

\citet{ha2013} have developed a phenomenological  model of a self-gravitating and accreting filament in which
turbulent support insures the filament stability. The turbulence is maintained by the kinetic energy of 
accreting  material, while the  dissipation comes from the ion-neutral friction. A key prediction 
of this model is that the thickness of the filament is indeed about 0.1 pc and importantly
does not depend on the density and column density of the filament. The reason stems from the fact that 
the ion-neutral drift operates on a timescale that is proportional to the ion density and that this latter
is proportional to the neutral density in this regime. This dependence cancels out with the square-root 
 of the gas density dependence of the freefall time.

However, so far this characteristic width has not been observed in numerical 
simulations. \citet{ntormousi2016} have performed a detailed analysis of the filament width distribution 
(in simulations that do not include self-gravity) and found 
that while ion-neutral friction affects the density structure and reduces the numbers of small scale filaments, 
it does not produce a characteristic width near 0.1 pc as can be seen in Fig.~\ref{eva}. 
This may be a consequence of the non-isotropic nature of this dissipation. 
In particular motions along the magnetic field lines are not dissipated by this mechanism. \\ \\

To conclude, let us stress that while some of these explanations succeed to explain the observed width in some specific range of column density,  none of the existing simulations performed so far have reproduced the characteristic width over 3 orders of magnitude in column density. 
Therefore, the origin of the apparent universal widths of the filamentary molecular clouds is still unclear. 
The problem could possibly be less severe because of the  bias due to finite resolution   
 \citep{gina2017} which may lead to artificial structures. 
Let us stress however that the massive filaments   are surrounded by an extended $r^{-2}$ 
envelope which has not been considered in the bias analysis of \citet{gina2017}, therefore these objects 
are clearly defined and apparently well resolved.
Note that it is quite possible that the bias described by \citet{gina2017} may also be present in the analysis  
of some of the numerical simulations.

\subsection{Fragmentation and core formation within filaments}
It has since long been recognized that cores often form in dense filaments 
\citep[e.g.][]{dutrey1991}
and several studies have performed stability analysis of hydrodynamical
\citep[e.g.][]{inutsuka1992} and 
magnetized filaments \citep[e.g.][]{nakamura1993,fiege2000,hanawa2015,hanawa2017}.
As the fastest growing mode has been found to be about four times  the filament diameter, 
\citet{inutsuka1992} argued that the fragments are expected to be separated by nearly 
four times this value. \citet{fiege2000} investigated the stability of filament 
threated by an helical magnetic field and conclude that although significant 
toroidal field can reduce significantly the growth rate of gravitationally driven 
modes, they lead to the development of the sausage instability.

Recent Herschel results have rejuvenated interest in filament forming cores.
In particular  \citep{poly2013,kony2015} have established that 
in nearby molecular clouds about 70-80$\%$ of dense cores lie within 
filaments. This may indicate that filaments are playing a significant role in the star formation process
although the mass distribution of cores lying inside and outside filaments may not be drastically different 
(see Fig.~17 of \cite{kony2015}). 

In light of recent results by Herschel, several other studies aiming at understanding the fragmentation of filaments in cores have been carried out to investigate various aspects of the non-linear fragmentation of filaments into cores. 
\citet{clarke2016} performed a series of numerical simulations to study the fragmentation of a filament that is accreting instead of being  at equilibrium as assumed in previous studies. 
Due to the gravo-acoustic modes induced by  accretion, 
the dispersion relation varies with  the accretion rate. 
\citet{mathias2017} carried out  simulations to study the response of a 
critical  filaments to bending modes. 
These modes, which tend to make the filament oscillates perpendicularly to its main axis, lead to  fragmentation.  The  cores which form  have a spacing that matches the wavelength of the sinusoidal perturbation of  the bending modes. 
Therefore  inferring filament 
properties from  characteristic spacing  should be considered  with care. 
\citet{clarke2017} performed  simulations where turbulence is seeded in accreting 
filaments and show that this generates fibers that are  similar to the ones
 observed in Taurus \citep{hacar2013}. 
They speculate that these fibers may  suppress radial collapse 
 within super-critical filaments.

Given the importance of filaments, it seems important to clarify the outcome of the fragmentation of filamentary molecular clouds and to understand the resulting properties of star forming cores. 
One of the most important outcome is the mass distribution of dense core, or so-called ``core mass function'' \citep{kony2015}.

\citet{chen2014} and \citet{chen2015} proposed a model for anisotropic core formation. 
In this model, filaments first form by flow of material along the magnetic field in post-shock layers 
where the field is strong. After filaments have acquired enough material that quasi-spherical 
regions are supercritical, strongly self-gravitating cores condense out. The two-step process predicts 
a characteristic core size and mass (Eq.~7 of \citet{chen2015}) and post-shock magnetic field that depends on the 
pre-shock density and inflow velocity but not on the pre-shock magnetic field strength.
Numerical results are generally consistent with this (see Fig. 10 and 11 of \citet{chen2015}).

The first attempt to obtain the mass function of prestellar cores from a 
filament structure
 was done by \citet{Inutsuka2001} in the case of the simple quasi-equilibrium filament, i.e., the filament supported by the thermal pressure and hence not radially collapsing. 
In particular, \citet{Inutsuka2001} found that
a line-mass spectrum $\delta^2 \propto k^{n}$ with $n\sim-1.5$ leads to
 a mass function of clumps  whose power law exponent is close to $-2.5$, i.e.,  $dn/dM \propto M^{-2.5}$. 
Note that the mass function discussed in his paper corresponds to the mass function of the systems, i.e. groups of stars, that may include binary or multiple stars. 
 \citet{Roy+2015} have recently measured the
 power spectrum of density fluctuations along  sub-critical filaments
 of the Gould Belt Survey. 
They  infer that $\delta^2 \propto k^{-1.6}$. 
If confirmed in a larger ensemble of filaments, this could explain the 
origin of the core mass function and its apparent 
universality.

\citet{lhc} have recently proposed an analytical theory to predict both the core mass function (CMF) and the mass function of groups of cores
of supercritical filaments.
The theory, which generalises the calculations performed by \citet{Inutsuka2001} and \citet{hc08}, considers magnetized filaments assumed to be radially supported 
by turbulent motions and takes into account thermal, turbulent and  magnetic supports.  
It predicts the CMF, which is found to depend on the mass per unit lengths (MpL) and the magnetic intensity. 
In particular, it is found that in the absence of magnetic field, filaments with  high MpL fragment in too many small cores. 
In the presence of magnetic field with moderate intensities and for sufficiently high MpL, CMF compatible with observed ones are inferred.

\section{The role of magnetic field in the evolution of molecular clouds and clusters}
In this section we more specifically address the role of the magnetic field regarding 
the evolution of molecular clouds 
as a whole and their ability to  form stars. We also discuss the properties of the 
star forming dense cores, which form in these clouds.

\subsection{Subcritical clouds}
Historically, one of the important questions related to the star formation 
process in the universe is the rate at which a galaxy is forming stars. 
In particular it is known since the work of \citet{zuckerman} 
\citep[see][for a more recent discussion]{kennicutt} that the star formation 
rate, at least in the Milky Way, is about hundred times lower than 
one would expect if the dense gas would be entirely in freefall. 
The origin of this factor hundred has remained mysterious 
during many years and magnetic field has been invoked to solve 
the problem \citep[e.g.][]{shu1987}.

To assess the importance of the magnetic field, one can compute the 
ratio of the magnetic over gravitational energies. 
As an illustrative example one can envisage 
a   uniform spherically symmetric cloud of mass $M$, volume $V$, radius $R$.
It is threaded by a uniform magnetic field of intensity $B$.  The magnetic
flux,  $\Phi$, is given by  $ \pi R^2 B$. 
In ideal MHD, the field is frozen into the gas and $\Phi$ remains 
constant. In this case we have
\begin{eqnarray}
{ E_ {\rm mag} \over E_{\rm grav}} = {B^2 V \over 8 \pi} \times {2  R \over 5 G M^2}
\propto {B^2 R^4 \over M^2} \propto \left( {\Phi \over M } \right)^2.
\label{emag}
\end{eqnarray}
Interestingly, $E_{mag} / E_{grav}$ is constant and in particular does not 
depend on the cloud radius. 

 It is clear from Eq.~(\ref{emag}), that
there is a critical value of the magnetic intensity for which the
gravitational collapse is impeded even if the cloud was strongly
compressed.  Mouschovias \& Spitzer (1976) have calculated accurately
the critical value of the mass-to-flux ratio using the virial theorem
and numerical calculations of the cloud bidimensional equilibrium.  A
cloud which has a mass-to-flux ratio smaller  than this
critical value cannot collapse and is called subcritical.  
It is called supercritical when the mass-to-flux is larger than the 
critical value.
It is usual to define $\mu = (M/\Phi) / (M/\Phi)
_{\rm crit}$. Large values of $\mu$ correspond to small magnetic
fields and thus supercritical clouds.

Considering a magnetically supported dense core, also called subcritical core, 
the evolution is considerably slown down being almost quasi-static for 
most of the time. The neutrals slowly cross the field lines and the magnetic flux 
is gradually reduced up to the point where the core becomes critical and 
dynamical collapse proceeds.
Estimating the time it takes is obviously the central question. 
To do so, clouds in virial equilibrium are considered, leading to
 $B ^2 / 4 \pi   \simeq M \rho G / R$.
The ratio of the ambipolar time, $\tau_{\rm ad}$ given by eq.~\ref{time}, and  the freefall time,  
$\tau_{\rm ff} \propto (G \rho)^{-1/2}$, 
\citep{shu1987} is then estimated to be
\begin{eqnarray}
{ \tau _{\rm ad} \over \tau_{\rm ff}} \propto {\gamma_{ad} C \over \sqrt{G}},
\label{time_ratio}
\end{eqnarray}
where it has been assumed that $\rho = C \sqrt{\rho_i}$. It is remarkable that 
in this expression there is no dependence in the physical parameters, such as the density, 
magnetic field and size. The exact value of  $\tau _{\rm ad} / \tau_{\rm ff}$ depends on 
the assumed geometrical coefficients. It is typically on the order of 10 (\citet{shu1987} estimated
$\tau _{\rm ad} / \tau_{\rm dyn} = 8$).

Equation~\ref{time_ratio} shows that the ambipolar diffusion process can reduce 
the star formation rate by almost an order of magnitude
if the field is strong enough to compensate gravity. 
This would bring the star formation rate much closer to the 
 observed values \citep[e.g.][]{shu1987}.  
To better quantify this process, 
one dimensional  simulations of subcritical clouds have been performed 
\citep[e.g.][]{basu1995}.
 For very subcritical  cores
 and  values of $\mu$ of about  $0.1$, \citet{basu1995}
inferred a collapse time equal to  15 freefall times. With critical cores,
 $\mu \simeq 1$, the collapse takes  roughly $\simeq$3 freefall
 times.

More recently, a series of simulations aiming at simulating subcritical 
and turbulent molecular clouds  have been performed 
\citep{heitsch2004, basu2004,linakam2004,nakamura2008,vanloo2008,nakamura2011,vazquez2011,bailey2014,bailey2017}.  
Typically it has been found that under the influence of  ambipolar diffusion but also
of turbulence, areas of high column densities and lower magnetisation develop. These
regions are typically supercritical and form gravitationally bound cores, that in turns 
form stars. This is portrayed in Fig.~\ref{naka08} where the column density in the direction 
of the initial magnetic field is shown as well as the isocontour of critical 
mass-to-flux ratio. 

In all these calculations, it has been found that subcritical magnetic fields 
decrease very substantially, down to few percent, the star formation rate. 
For example Fig.~\ref{vazquez} shows the mass of the dense gas and the mass within 
sink particles for a series of calculations including different magnetizations
and with or without ambipolar diffusion. As can be seen the mass within 
sink particles (i.e. ``stars'') is almost 2 orders of magnitude smaller with 
an initial magnetic field of 4 $\mu$G than with a field of 2$\mu$G (one must keep 
in mind that these values correspond to the magnetization of the diffuse gas 
out of which the molecular cloud is assembled). It is also interesting to notice 
that in these calculations, the ambipolar diffusion makes only a modest 
difference. This indicates that a lot of magnetic flux is actually diffused 
through turbulence rather than ambipolar diffusion. Alternatively, this may also 
indicate that some gas is being accreted along the field lines, therefore reducing 
locally the mass-to-flux ratio.

Note that the interaction between turbulence and ambipolar 
diffusion can be complex. For example \citet{linakam2004} and \citet{nakamura2008} 
found that stars 
may actually form more rapidly when the turbulence is higher because turbulence leads 
to a faster ambipolar diffusion by creating stronger shocks where the gradients of 
magnetic field are steep.

\setlength{\unitlength}{1cm}
\begin{figure*} 
\begin{picture} (0,10)
\put(0,0){\includegraphics[width=10cm]{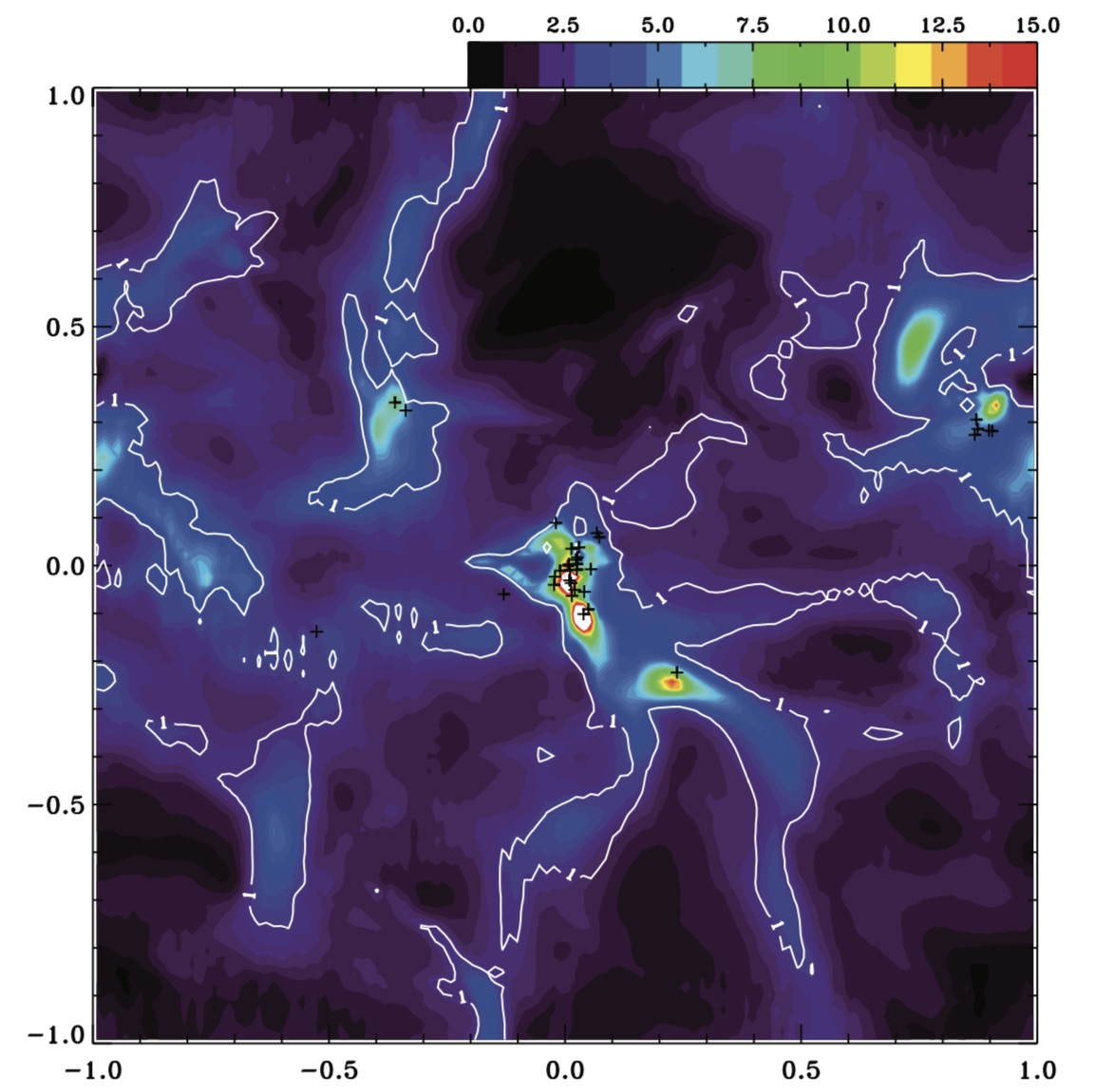}}
\end{picture}
\caption{Spatial distribution of column density and mass-to-flux ratio in an initially subcritical
and turbulent  molecular cloud
 \citep{nakamura2008}.  Inside the contour the gas is supercritical and has collapsed at several places. 
Outside the contour it is still subcritical.
Reproduced from \citet{nakamura2008} with permission of ApJ.}  
\label{naka08}
\end{figure*}

\setlength{\unitlength}{1cm}
\begin{figure*} 
\begin{picture} (0,6)
\put(0,0){\includegraphics[width=7cm]{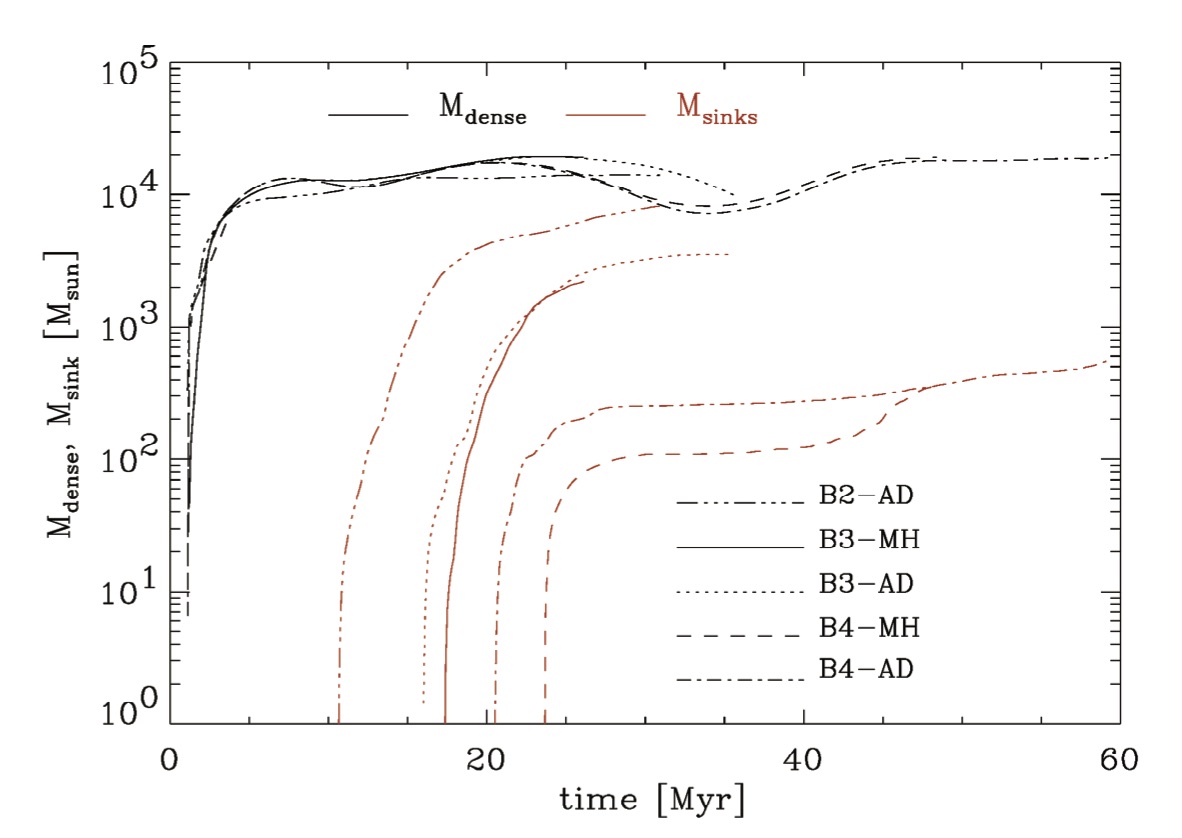}}
\put(7,0){\includegraphics[width=12cm]{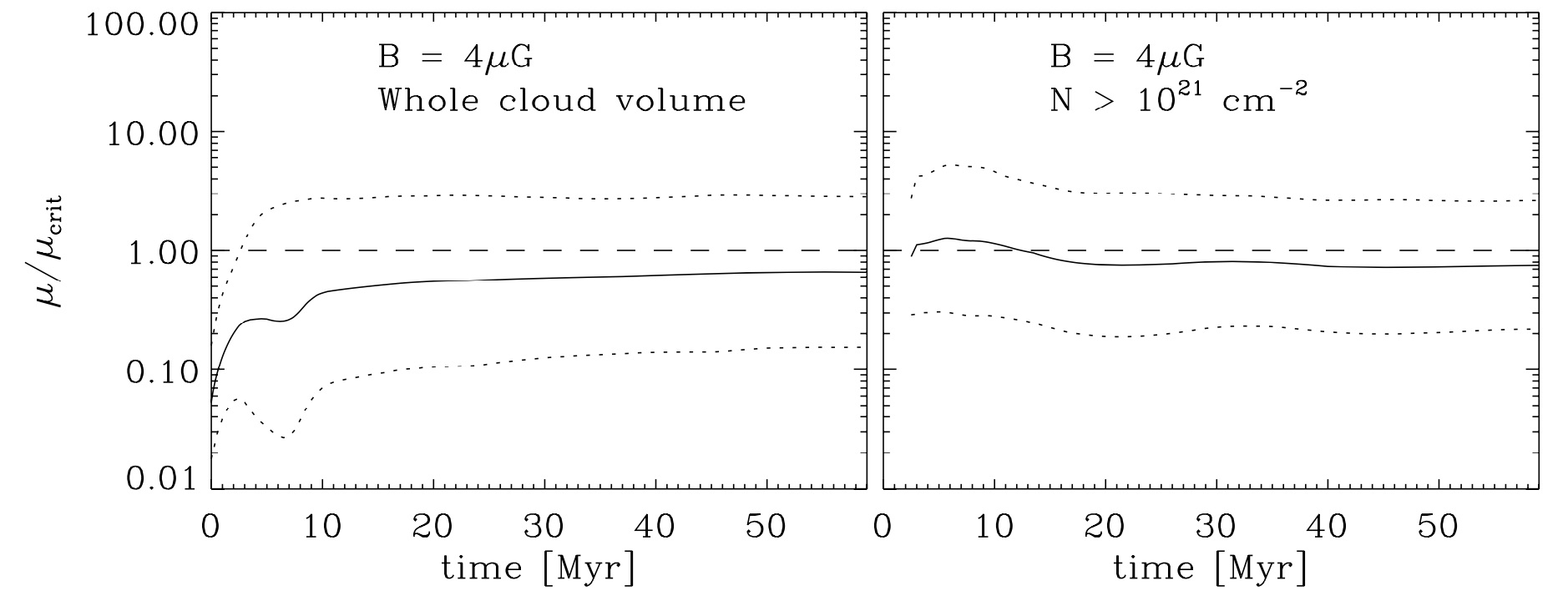}}
\end{picture}
\caption{Mass of dense gas and distribution of mass-to-flux ratio in 
a molecular clouds formed from colliding flows
\citep{vazquez2011}. Three initial magnetizations are being shown, 
namely 2, 3 and 4$\mu$G and runs with and without ambipolar diffusion have been performed.}
\label{vazquez}
\end{figure*}

\subsection{The properties of cores in magnetized molecular clouds}
We now turn to a discussion on the core properties that have been inferred from 
MHD simulations. As various rather different setups have been inferred we first 
start with a brief description of the numerical experiments, which 
have been performed. Here we restrict the discussion to studies that 
have explicitly discussed core statistics, i.e. self-gravitating structures
formed within  simulations that handle MHD and gravity. The cores are 
identified using a clump finding algorithm and then only the ones which are effectively 
self-gravitating are selected. Given that the typical size 
of dense cores is a fraction of 0.1 pc \citep[e.g.][]{ward-thompson2007}, the 
spatial resolution achieved in these calculations is typically a fraction of 
0.01 pc.

\subsubsection{Numerical setups of the various numerical experiments}
As several setups and initial conditions have been considered, we first give a quick overview 
of the different choices that have been made. 

\paragraph{Prescribed molecular clouds}
Many numerical experiments start with a uniform density cloud or a mildly 
peaked one \citep{basu2004,linakam2004,vazquez2005,tilley2007,nakamura2008,nakamura2011,bailey2014,bailey2017}.  
The gas is assumed to be isothermal and self-gravity is treated. 
The initial field is usually uniform and various intensities ranging from 0 to significantly
magnetized, are being assumed.
 Some velocity field with an amplitude corresponding 
to a Mach number up to 10, is usually prescribed. In these works either
artificial driving of the turbulence is applied \citep{vazquez2005} either turbulence is free to decay. 
\citet{nakamura2008} and \citet{nakamura2011} include protostellar outflows, which drive turbulent motions.
Ambipolar diffusion in the strong coupling limit is applied in some of these works.

\paragraph{Colliding flows}
The colliding flow setup \citep[e.g.][]{hetal2008,baner2009,vazquez2011,clark2012,valdivia2016} has also been used to study 
core formation \citep{chen2014,chen2015}.
It consists in imposing two streams of gas with supersonic sound speed that create a dense
shocked layer, which eventually gives rise to a layer of denser gas. The advantage of this setup is that the cloud is 
built and not imposed as it is the case for the previous setup. In particular, it is initially not self-gravitating. 
The other advantage is that the turbulence within the cloud is a consequence of the incoming flow. 
\citet{chen2014} and \citet{chen2015} vary the Mach number of the incoming flow and the magnetization. They treat 
ambipolar diffusion in the strong coupling approximation. In this scenario the transverse component of 
the magnetic field is amplified in the shock bounded layer as the colliding flow is super-Alfv\'enic.

\paragraph{Zooming-in from galactic box calculations}
One of the restrictions of the two previous setups is that the initial conditions or boundary conditions have 
to be assumed. Moreover the statistics remain limited because, to unsure good resolution, the computational 
domain is typically few parsecs across. To circumvent these difficulties, \citet{h2018} has performed 
adaptive mesh refinement simulations of a kpc numerical domain. These simulations includes  stratification, induced by 
the gravitational field due to stars and dark matter. They start with only WNM and have an initial magnetic field 
parallel to the equatorial plane of about 3$\mu G$. In a first phase, supernova driving is applied. Once a self-consistent 
multi-phase and turbulent ISM is obtained, nine levels of adaptive mesh are employed to refine a region of 
100$\times$100 pc$^2$. This provides a final resolution of about 4$\times 10^{-3}$ pc.

\subsubsection{Core properties}

\setlength{\unitlength}{1cm}
\begin{figure} 
\begin{picture} (0,11)
\put(5.1,0){\includegraphics[width=6.5cm]{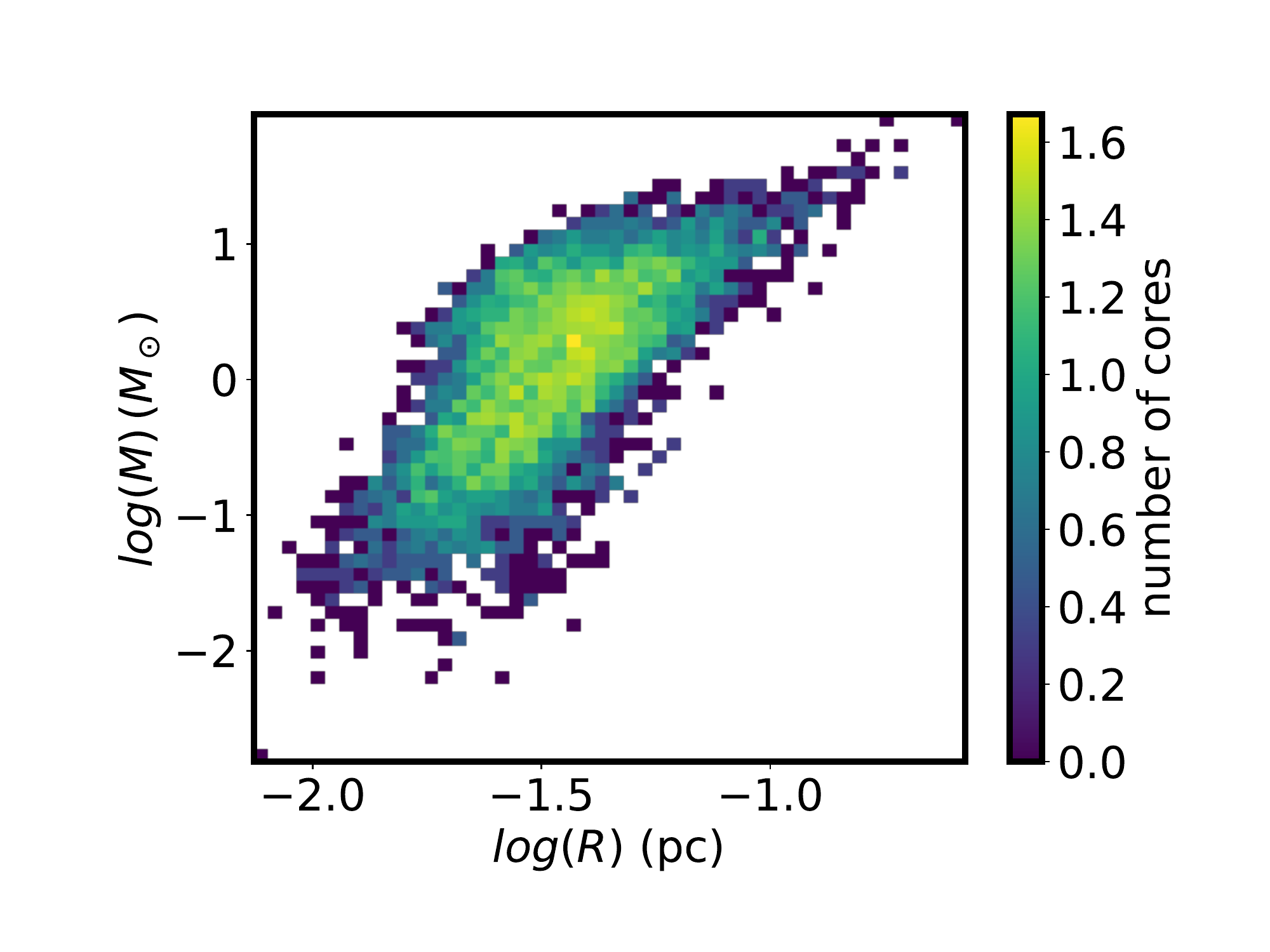}}  
\put(-1,0){\includegraphics[width=6.5cm]{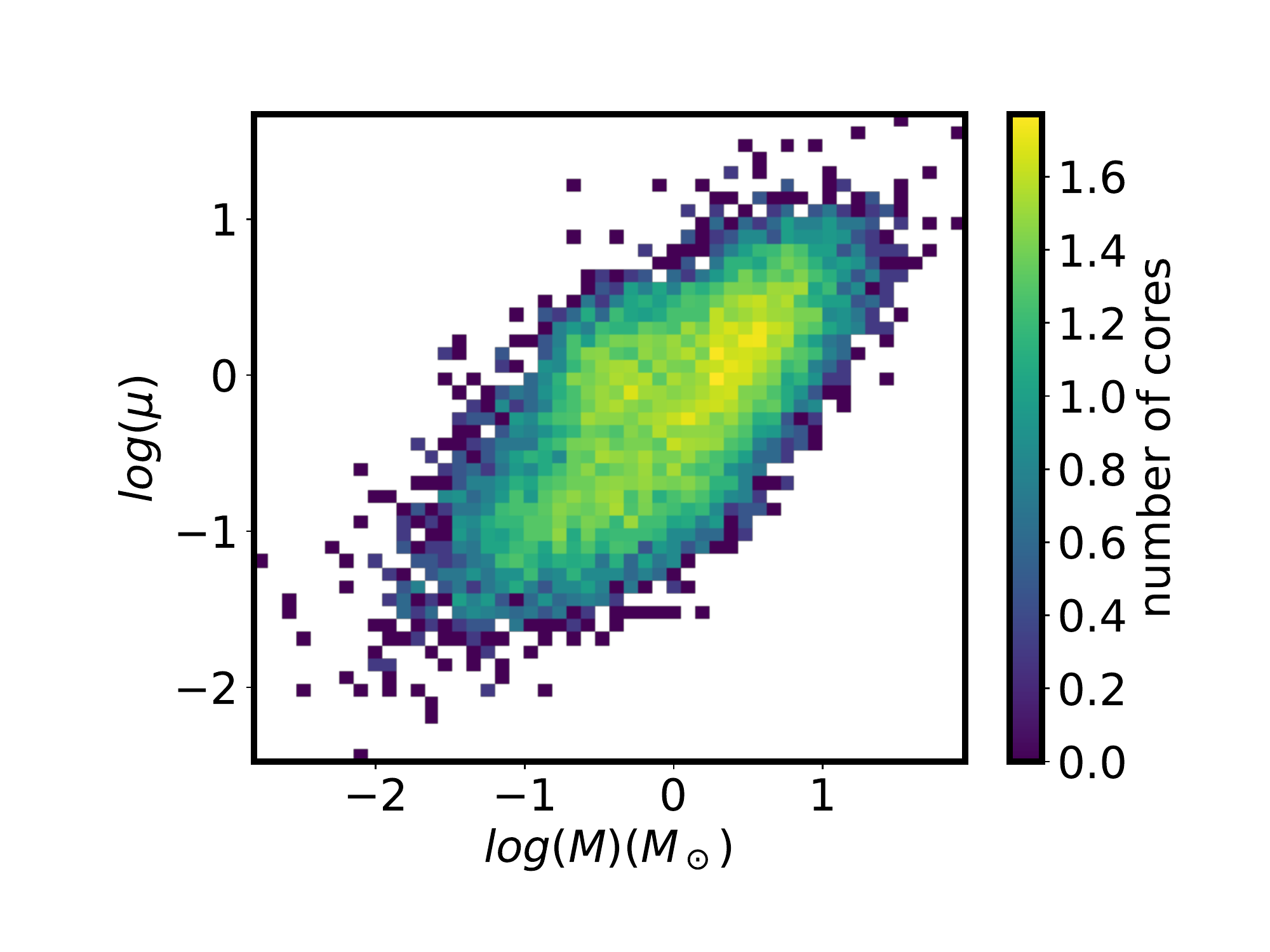}}  
\put(11.5,0){\includegraphics[width=6.5cm]{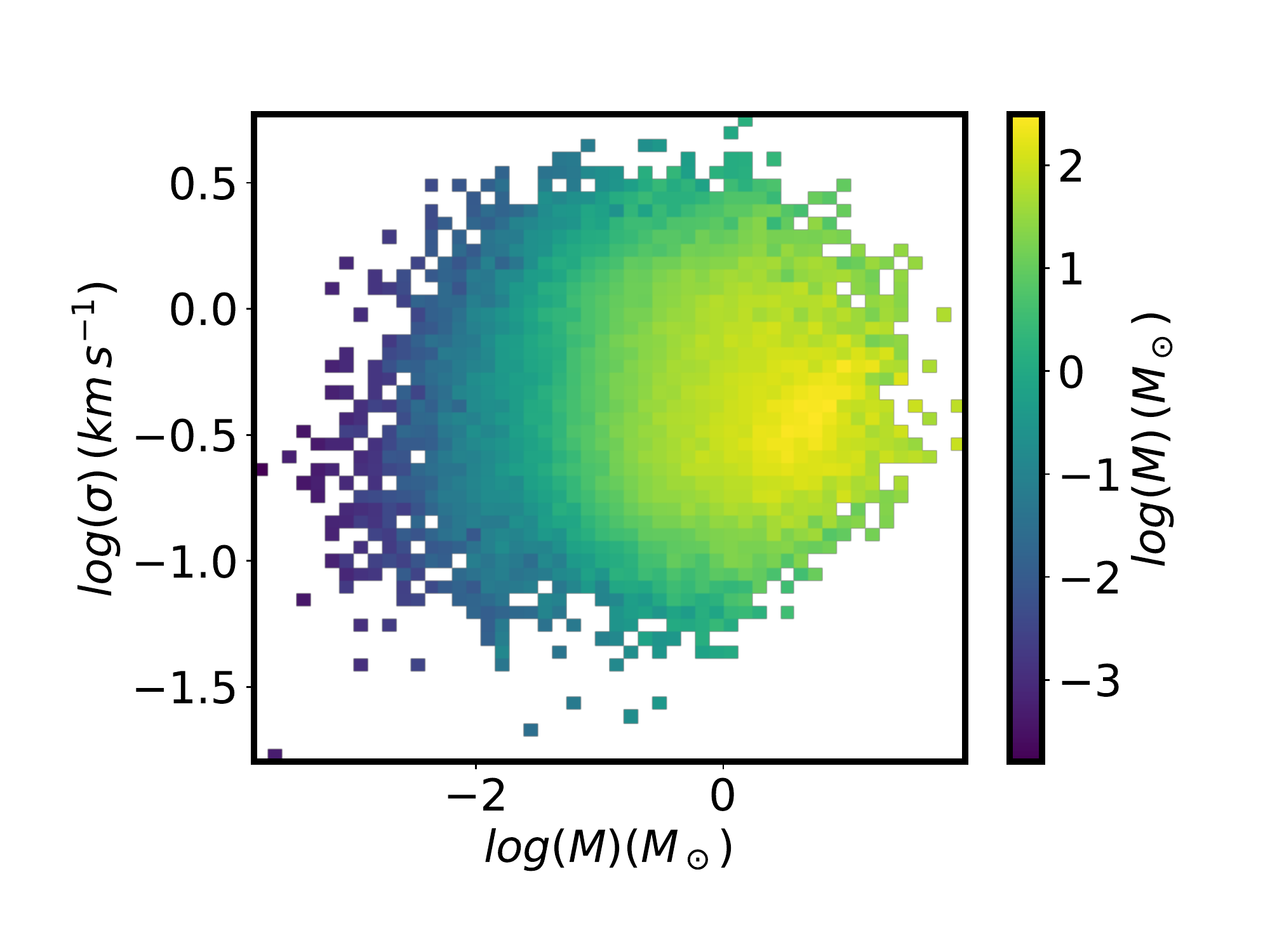}}  
\put(11.5,5){\includegraphics[width=6.5cm]{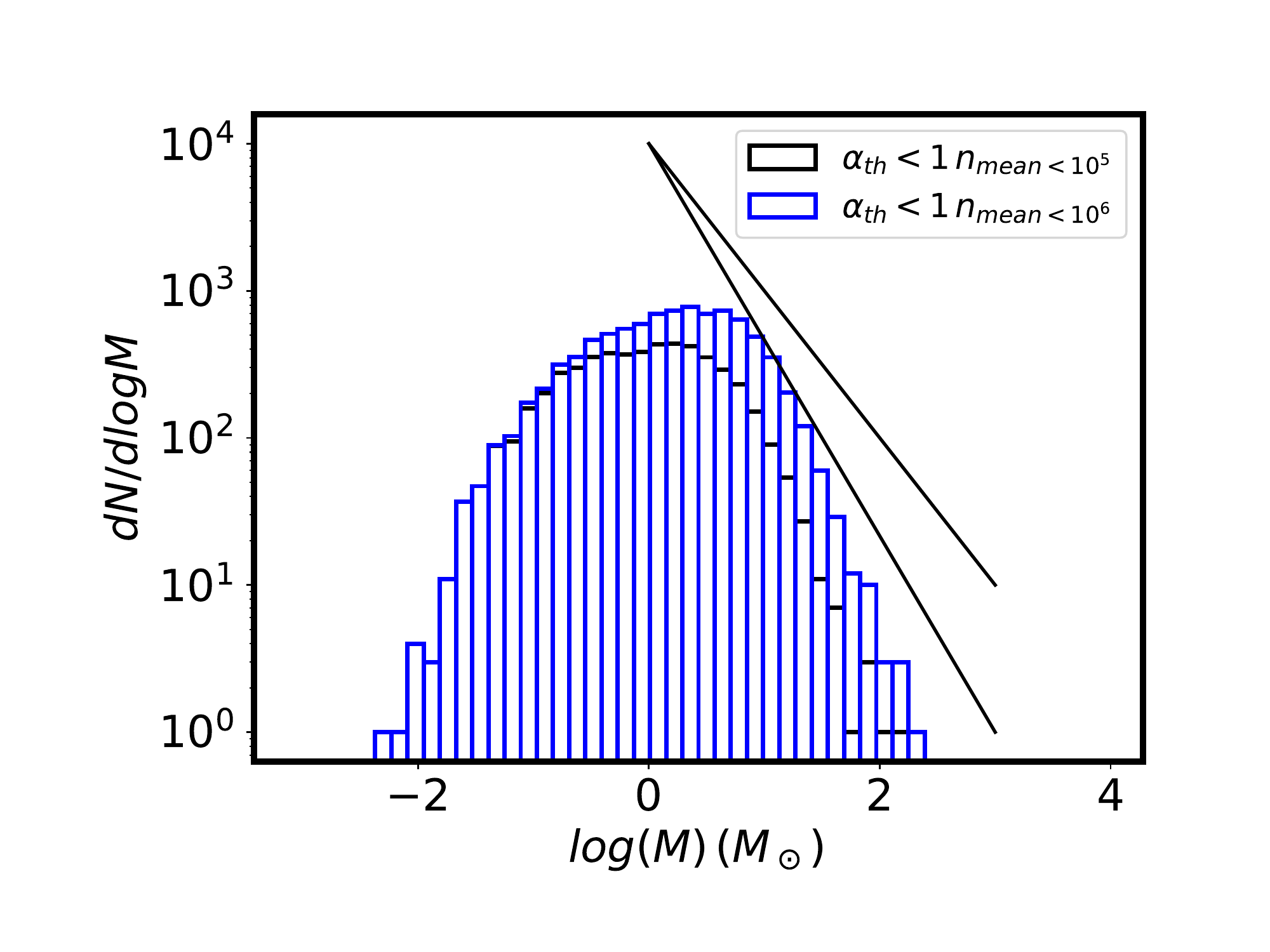}}  
\put(5,5){\includegraphics[width=6.5cm]{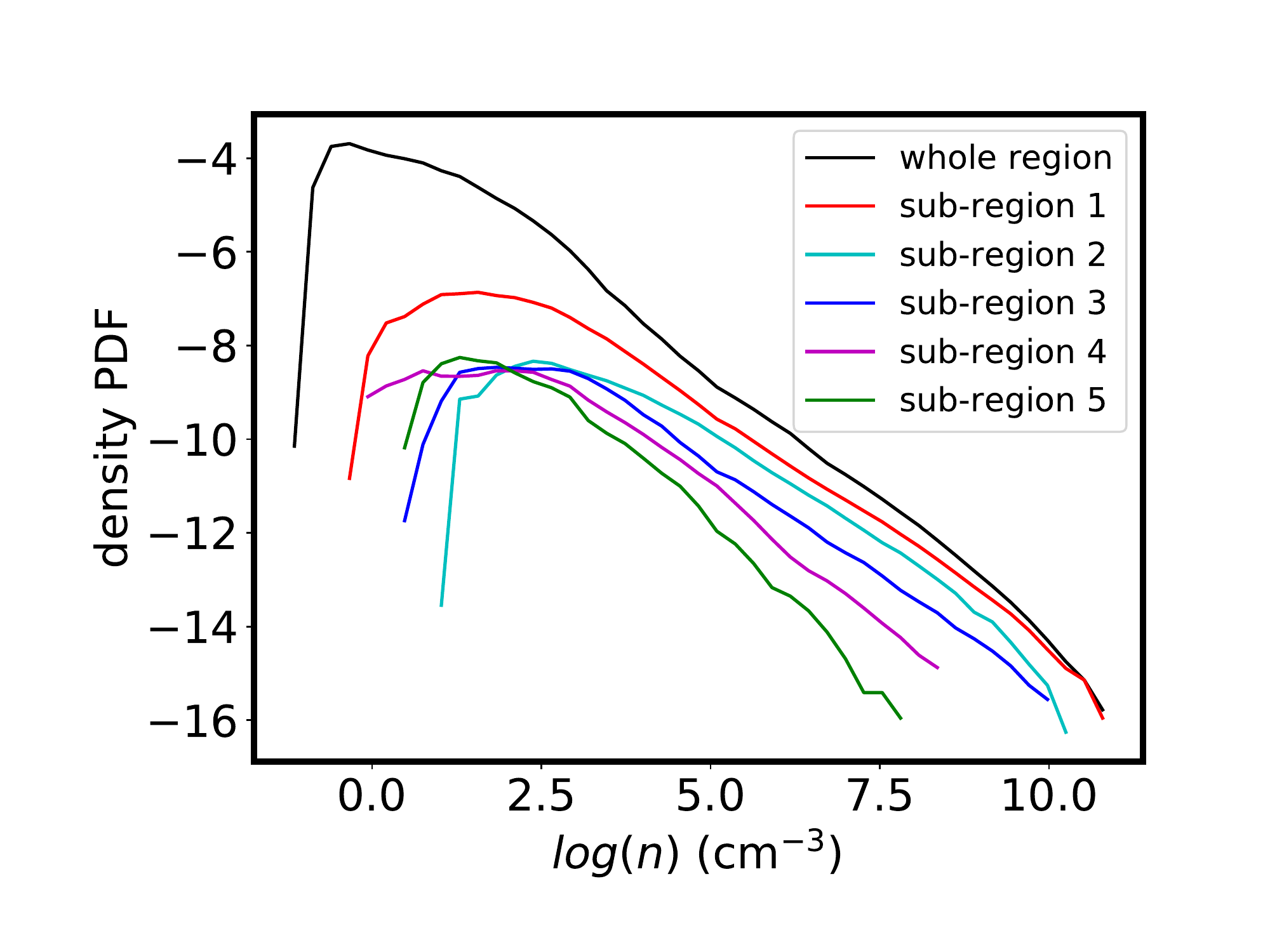}}  
\put(-1,5){\includegraphics[width=6.5cm]{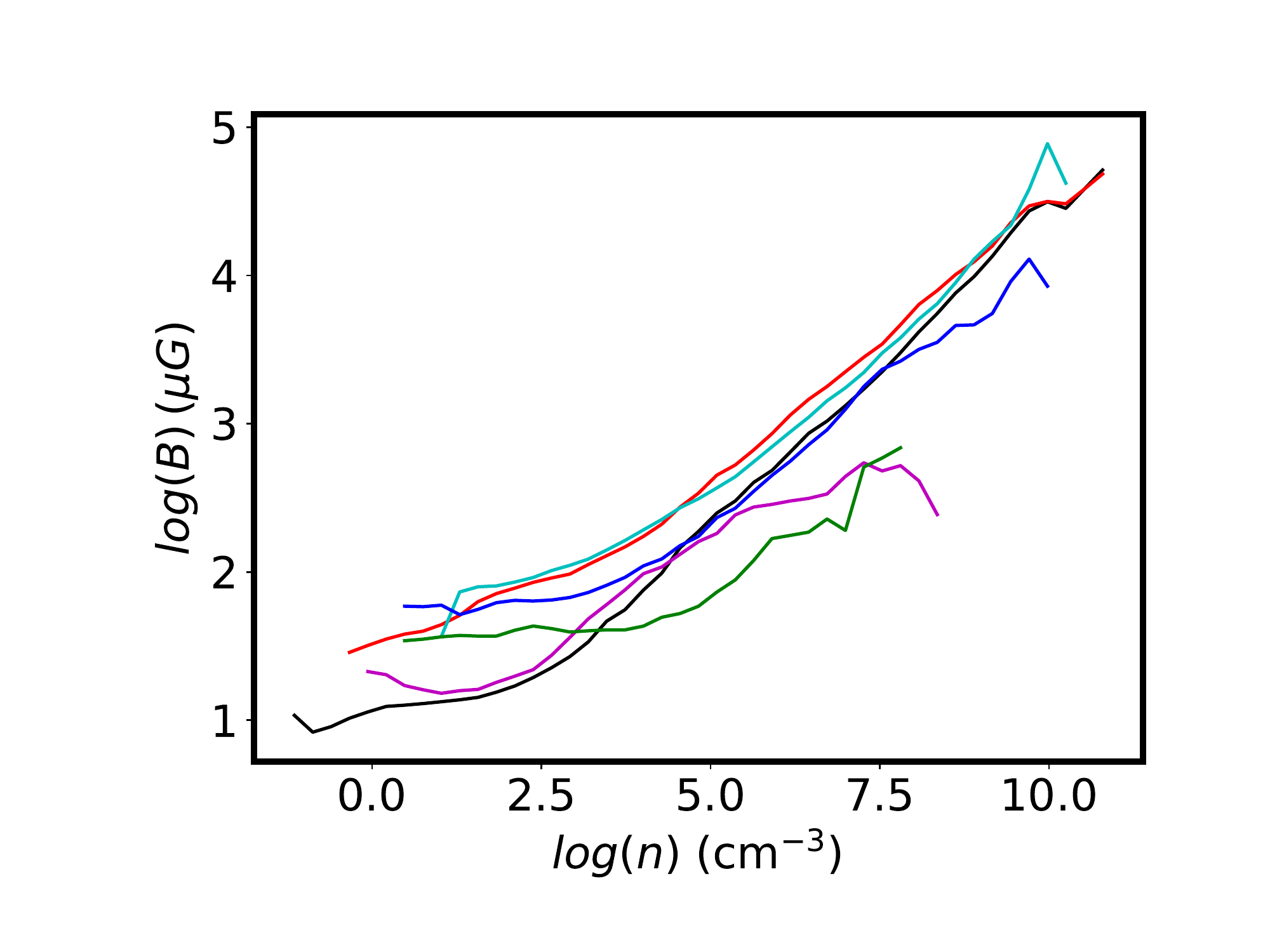}}  
\end{picture}
\caption{Global flow properties and dense core statistics in a zooming-in calculation
that goes from 1 kpc box size to $4 \times 10^{-3}$ pc resolution \citep{h2018}.
 Top-left panel: magnetic intensity vs density in the whole refined region of the ARM simulation and in 5 star-forming subregions.
Top-middle panel: same as top-left panel for the density distribution.
Top-right: mass spectrum of dense cores (defined as thermally supercritical objects for 2 different thresholds of their mean 
density).
Bottom-left panel: mass-to-flux ratio vs mass for the thermally supercritical cores.
Bottom-middle: mass-size relation of  the thermally supercritical cores.
Bottom-right: internal velocity dispersion relation vs mass of all cores.
 Reproduced from \citet{h2018} with permission of A\&A.} 
\label{pdf_clust}
\end{figure}

In the studies presented above, many core properties have been inferred. Here we restrict the discussion to 4 of them,
comparing the results obtained in the different configurations explored. 
In spite of the relatively broad diversity of these latter, the results are in relatively good agreement.

\paragraph{Core mass spectrum}
The mass spectrum of dense cores is likely to be important because they constitute the eventual mass reservoirs
of stars. Moreover the core mass function (CMF) has been found to have a shape similar to the stellar initial mass
function (IMF) \citep{ward-thompson2007,offner2014}, seemingly suggesting that the CMF may be at the origin of the IMF. 
Note that this possible link between the CMF and the IMF is still a matter of debate and numerous
studies argue that the IMF is not linked to the CMF (see Section~\ref{frag_clust} and \citet{offner2014} for a
recent review on this topic).
The CMF has been computed by \citet{tilley2007}, \citet{nakamura2008}, \citet{nakamura2011}, \citet{chen2014} and \citet{h2018}.
It has generally been found that the CMF {\it resembles} the observed ones \citep{kony2015}. In particular, 
it presents a peak and  a powerlaw at large masses (see top-right panel of Fig.~\ref{pdf_clust}) 
with a slope close to the observed one 
\citep{tilley2007,nakamura2011,h2018}. The slope is compatible with the idea that cores form under the combined 
influence of gravity and turbulent support \citep{hc08} while magnetic field does not appear to have a 
strong influence \citep[e.g. Fig.11 of][]{nakamura2011} in good agreement with theory  \citep{hc2013}. 

The question of the peak is far less clear. Observationally a peak around 0.5-1 $M_\odot$ has been inferred
\citep{kony2015}, though higher resolution observations need to confirm its robustness. 
In the simulations the existence of the peak must also be handled with care. First of all, simulations that 
have no preferred scales like isothermal simulations with ideal MHD can be freely rescaled to any units. This 
means that the peak is a direct function of the initial conditions. Second of all numerical convergence 
must be carefully verified. For example \citet{h2018} performed runs at different resolutions and 
concluded that indeed the peak of the CMF varies with numerical resolution.  Note that \citet{gong2015}
on the contrary concluded that numerical convergence is reached in their colliding flow calculations.
The most likely reason of this discrepancy comes from the differences of the physical conditions studied
and in particular the global gravitational stability of the simulated regions. Indeed gravity 
induces density PDF with high density powerlaw tails in which case the CMF may not present a 
peak at all (see discussion in \citet{lee2018a} and \citet{lee2018b}).

Once rescaled to the mean Jeans mass \citep[e.g.][]{tilley2007,chen2014,chen2015}, 
the dependence of the CMF on physical parameters has been found to be  limited. In particular, 
 \citet{chen2015} found a modest dependence of the CMF onto the magnetic intensity. This is at first 
surprising as magnetic field is part of the total support. To account for this weak 
dependence, they propose an anisotropic scenario in which contraction first start along the field lines before 
enough mass is accumulated to trigger contraction perpendicularly to the field lines. 

A complementary information is provided by the mass-size relation (displayed in bottom-middle panel
of Fig.\ref{pdf_clust}) which has been studied by \citet{chen2014}, \citet{chen2015} and \citet{h2018}. 
Typically a relation $M \propto R^\alpha$, with $\alpha \simeq 2$ has been inferred.

\paragraph{Magnetization}
The magnetization of cores is a fundamental parameter to determine.
Indeed magnetic field 
has been found to influence significantly the collapse of 
cores and in particular the formation of planet-forming disks through 
magnetic braking \citep{inutsuka2012,lietal2014,hetal2016}. 

\citet{chen2015} provide (Fig.12) the mass-to-flux as
a function of the core mass. They found that most cores
are supercritical with typical values for $\mu$ of about
 2. A clear trend is seen for $\mu$ to increase with the 
core mass. The dependence of the $\mu$ distribution on the 
initial large scale magnetic field and the Mach number 
of the colliding flow
is found to be relatively weak 

\citet{h2018} has been measuring the mass-to-flux ratio in cores
identified as thermally supercritical, that is to say cores that 
would collapse if only thermal support was present. 
The result 
is displayed in bottom-left panel of Fig.\ref{pdf_clust} where 
the mass-to-flux is displayed as a function of the core mass. 
The mass-to-flux is found to increase with the mass and is 
roughly proportional to it with $\mu \simeq 1$ for $M \simeq 1 \, M_\odot$.
However, a relatively broad distribution 
is inferred and for a given mass, the mass-to-flux distribution spans
almost one order of magnitude.
Note that many of these cores would actually not collapse (unless they accrete 
more mass along the field lines) as they are subcritical and therefore 
magnetically supported.

\paragraph{Velocity dispersion}

The velocity dispersion in and around cores has received 
a lot of attention. \citet{nakamura2008} and 
\citet{nakamura2011} found that the velocity 
dispersion in cores present a large spread and goes
from sonic (i.e. the velocity dispersion is close to the 
sound speed) to a Mach number larger than 5, 
with values up to 2-3 km s$^{-1}$.  They found that 
there is no clear dependence of the velocity dispersion with the 
mass or the size. This is very similar with what is reported
in \citet{h2018} and displayed in the bottom-right panel of 
Fig.\ref{pdf_clust}. 

\citet{nakamura2011} reported that the velocity dispersion 
is significantly reduced when the magnetization is high. Typically
the cores formed in highly magnetized clouds tend to have trans-sonic to 
sub-sonic motions only.

\paragraph{Magnetic field orientation and core shape}
The relationship between core shapes and the direction of the magnetic field has been analysed by 
 \citet{chen2018}. They found that in colliding flow simulations  cores are generally triaxial, 
and the magnetic field tends to be parallel to the shortest axis and perpendicular to the longest axis, 
with internal and external magnetic field direction correlated. 
This is a natural consequence of the formation of cores within filaments and the 
fact that magnetic field tends to be perpendicular to self-gravitating filaments as explained previously
\citep[e.g.][]{soler2013,gomez2018}. 
They also found that
 core angular momentum vectors are not aligned 
with the direction of the (internal or external) magnetic field.
As explained below, this may be important in the context of protoplanetary disk formation.

\subsection{The influence of magnetic field on low mass collapsing cores}
The collapse of low mass prestellar cores leads to the formation of small groups of stars, a process
known as fragmentation, and to the formation of protoplanetary centrifugally supported disks. 
It is presently believed that magnetic field has a drastic influence on the outcome of 
collapsing cores. 

\subsubsection{The magnetic braking process}
A fundamental difference between hydrodynamical and magnetized prestellar cores comes 
from the evolution of angular momentum. In the absence of a substantial magnetic field, 
the latter is essentially conserved and becomes dominant drastically affecting 
the evolution of the core \citep[e.g.][]{matsumoto2003}. 
 In a magnetized core, the situation is different. Because of magnetic tension, angular momentum 
can be exchanged between  fluid particles. Typically this exchange occurs between  a cloud and an 
intercloud medium and happens through 
torsional Alfv\'en waves, which propagate in the  intercloud medium 
\citep{mouschovias1981,shu1987,joos2012}).
To estimate the characteristic time scale  of magnetic braking 
let  $\rho_{\rm icm}$ be the density of the intercloud medium.  For simplicity, we consider that
 the magnetic field is parallel to 
the cloud rotation axis. The torsional Alfv\'en waves propagate  at a speed, 
$V_a= B / \sqrt{4 \pi \rho_{\rm icm}}$.
The magnetic braking is important if  a significant fraction of the 
cloud angular momentum has been delivered to the intercloud medium. 
 This occurs  when the waves have reached a distance  
 $l \times \rho_{\rm icm} \simeq R \times \rho_0$. 
This leads to
\begin{eqnarray}
\tau _{\rm br} \simeq {R \over V_a} {\rho_0 \over \rho_{\rm icm}}. 
\label{brake}
\end{eqnarray}
Equation~(\ref{brake}) is obtained assuming a very simple geometry. Other estimates in different
geometries can be found in the references mentioned above. In particular, the braking depends
on the angle between the magnetic field and the rotation axis, it also depends whether the field 
lines are uniform or fan out, in which case the braking time can be considerably reduced.

\subsubsection{Disk formation: a magnetically controlled process ?}
In the aligned configuration, the braking time can become so short that the formation of the 
centrifugally supported  disks can be even entirely prevented 
\citep[e.g.][]{allen2003,galli2006,price2007,hf2008,lietal2014}, 
a process known as catastrophic braking. More recent studies have revealed that the aligned configuration is 
however too simplified and that disks should form in magnetized clouds, although in general the disks are 
smaller and fragment less than in the hydrodynamical case. These studies fall in two categories. First, 
the magnetic braking is reduced because i) the magnetic field and the rotation axis are non-aligned 
\citep{joos2012,gray2018}, ii) the turbulent velocity field diffuses the magnetic field 
\citep{santos2012,joos2013}, iii) the turbulent field makes the structure of the magnetic field 
less coherent \citep{seifried2013}. Note that \citet{gray2018} performed turbulent simulations in which 
the angular momentum is aligned with the magnetic field and show that disks do not form or are much smaller 
than in the same simulations for which there is no alignment. They concluded that misalignment may be the dominant effect.
The second category of processes that limits catastrophic braking is non-ideal MHD. This has been studied by 
numerous groups \citep{inutsuka2012,lietal2014,hetal2016,masson2016,machida2016,wurster2016,zhao2018}.
These works found that small disks (i.e. disks significantly smaller than in the hydrodynamical case) form. 
Moreover, the turbulence and the magnetic configurations (i.e. misalignment) tend to be less important when 
non-ideal MHD processes are accounted for.

\subsubsection{How magnetic field changes the fragmentation of low mass cores}
In typical conditions, that is to say with low rotation speed and relatively high thermal 
support, low mass cores generally fragment in several objects \citep{matsumoto2003}.
This fragmentation is due to the generation of the density fluctuations induced by 
turbulence or by gravity itself. Rotation considerably helps by maintaining important quantities 
of gas in  equilibrium leading to the formation of massive, highly unstable disks. The issue of 
fragmentation is therefore strongly dependent of the initial conditions, namely the rotation and turbulence 
level as well as the presence of density perturbations initially. 
Several studies have been dedicated to the influence of magnetic field in this process
\citep[e.g.][]{machida2005,machida2008,ht2008,commercon2010,wurster2017}. 
It has been found that when the density perturbations are low, typically 10$\%$ or so, 
the magnetic field is drastically reducing fragmentation, which happens only 
when the magnetic intensity is low. This is because when no strong 
density perturbations is initially present, the fragmentation occurs through rotation and 
the formation of massive, highly unstable disks. However, magnetic field gets efficiently wind up
by the differential rotation which develops in the core inner part. As discussed 
above, angular momentum is then efficiently extracted and the disks are smaller. 
Another important effect, which further reduces rotationally induced fragmentation, is 
the magnetic pressure itself, particularly the one associated to the toroidal 
magnetic field. This pressure adds up to the thermal one and makes the disks 
more stable \citep{ht2008}. This stabilisation seems to persist even  when non-ideal
MHD effects are accounted for \citep{wurster2017}.  
On the other hand  density perturbations of large amplitude, that is to say  of 
about 50$\%$,  are sufficiently unstable to collapse individually even in the absence of 
rotation. In this case, magnetic field is unable to impede fragmentation \citep{ht2008,wurster2017}.

\subsection{Is magnetic field playing a role in the formation of clusters ?}
It is believed that stars do not form in isolation but rather in clusters \citep[e.g.][]{longmore2014}. 
Indeed, observationally stars do not form in the bulk of molecular clouds but instead 
in their denser parts. 
Large surveys have recently revealed  ensemble of massive clumps in which 
stars are actively forming \citep{fall+2010,urquhart+2014,traficante}. These clumps
have masses up to several thousands of solar masses and  are very good candidates 
for being stellar cluster progenitors. 
Given that these objects are relatively massive it is unlikely that magnetic field 
plays a major contribution on the formation and global equilibrium 
of these massive clumps. Numerical simulations are able to reproduce reasonably well the 
global properties of these massive clumps such as their mass-size relation, 
simply by invoking gravity and turbulence \citep{lee2016a,lee2016b},  starting with reasonable 
ISM magnetic intensities, 
magnetic energies a few times above the thermal ones but well below the kinetic 
and gravitational energies are obtained.

There are however several aspects that deserve particular attention and which are now examined. 
First of all,  
does  magnetic field influences the small scale fragmentation, that is to say the formation of 
the stars themselves ? 
Does magnetic field enhance   stellar feedback ? 
Does  magnetic field increase the coupling between clusters and outflows and jets ?

\subsubsection{Does magnetic field affect the small scale fragmentation in stellar clusters?}
\label{frag_clust}
One fundamental goal of cluster studies is to infer the mass function of stars that form and 
whether it can reproduce the IMF. Several studies have been investigating this issue
using sink particles \citep[e.g.][]{bate2012,myers2013,offner2014,lee2018a}. As the present review focusses 
on the possible  role of magnetic field,  the discussion below is restricted to this aspect specifically.

\paragraph{MHD barotropic calculations}
Before we describe the calculations that include radiative 
processes, we first consider the simpler 
barotropic case. 
Such simulations  have been performed 
by  \citet{price2008} and \citet{hetal2011} 
\citep[see also][who even included photo-ionisation from the central star]{peters2011} 
who have simulated the collapse of
several tens  $M _\odot$ turbulent cores. The  turbulent and gravitational
energies were initially comparable and various magnetic intensities
have been explored. It has been found that 
the fragmentation is reduced when the  mass-to-flux is smaller than $\simeq$5.
The fragment number is reduced by roughly a factor of 2 
for the strongest magnetisation.
It has been found that during the collapse, efficient 
magnetic diffusion occurs due to the turbulent 
velocity field, which explains why fragmentation is reduced by a
factor 2 only.

Recently \citet{lee2019} studied the collapse of 1000 $M_\odot$ clumps
with various magnetization. As sink particles, are being used, the initial 
mass function was studied. They concluded that while magnetic field reduces
a bit the number of objects,  it is of secondary 
importance to determine the  shape of the IMF. This is in part due to the fact that the peak 
is determined at very small scale (100 AU or so) and at very high density where the magnetic intensity 
weakly depends on the large scale initial conditions and therefore tends to have 
the same value irrespectively of the initial value.

\paragraph{MHD radiative calculations}
Collapse calculations of massive collapsing cores, in which 
both the magnetic field and the radiative transfer have been
taken into account, have been carried out.

 \citet{pricebate2009}, found  that the 
magnetic field and the radiative feedback play  complementary effects. 
Magnetic field  supports the diffuse gas at large scale and radiative feedback, 
by  heating the inner part of the core, reduces the  fragmentation in many objects.

Simulations including magnetic field and radiative feedback,
which  follow the collapse up to AU scales, have been 
performed by \citet{commercon2011}, \citet{myers2013}, \citet{myers2014}
and \citet{cunningham2018}. 
It has been found  that in some circumstances, 
the combination of magnetic field and radiative feedback 
may be reducing fragmentation significantly.
This is due to the fact that 
magnetic field induces 
 efficient magnetic braking and  reduces the 
amount of angular momentum in  the cloud inner part. 
Consequently, the accretion is 
 more focused in a magnetized core 
than in an hydrodynamical one.  In this latter case
strong angular momentum prevents the gas to fall in the cloud
center. Thus the accretion luminosity, which 
is $\propto M \dot{M} / R$ is much higher since $M$
 and  $\dot{M}$  are larger while $R$ is smaller.
The temperature in magnetized cores is therefore higher 
than in hydrodynamical ones and this reduces the fragmentation within 
the former.

\subsubsection{Does magnetic field enhance  stellar feedback ? }
A possibly important consequence of the magnetic field could be 
related to this very last point. This is because 
stellar feedback critically depends on the stellar masses. 
This is the case for the HII radiation, the winds and 
of course the supernova explosions, which require the 
stellar mass to be larger than 8 $M_\odot$. 
For example numerous authors \citep{dale2011,walch2012,geen2015,geen2017}
found that in Milky Way type conditions, HII regions likely destroy molecular clouds  
quickly after they form stars, likely limiting the star formation efficiency of these objects. 

\citet{arthur2011} 
\citep[see also][]{mackey2011,gendelev2012} performed both unmagnetized and magnetized simulations of the expansion of an HII 
region in a molecular clumps and studied in details the resulting structure of the field. They found
that the magnetic field does not change very significantly the expansion in itself but reduces 
the small scale fragmentation and radiation-driven  pillars. The field in the neutral expanding shell
is preferentially parallel to the shell while in the ionised gas inside the shell it is 
more perpendicular to it. 

Since magnetic field tends to reduce fragmentation, 
it is likely that without magnetic field the stars would be on average less massive and therefore
their HII radiation which is proportional to $M^{2-3}$ \citep[e.g.][]{vacca1996}, would be 
significantly reduced. 
Since numerical simulations are not able yet to self-consistently predict the mass of the stars 
and follow the large scale evolution of the parent clouds subject to their feedback, it is 
however not possible to get a firm confirmation of this effect.

\subsubsection{Does magnetic field improve the coupling  with jets ? }
The influence that jets may have on a proto stellar clusters
has been investigated  at pc scales
\citep{linakam2006,cunningham2009,carroll2009,wang2010,fed2015} 
and inside massive cores \citep[e.g.][]{cunningham2011}.

 \citet{linakam2006}  \citep[see also][]{wang2010,fed2015}
carried out calculations for  a 10$^3$ M$_\odot$ clump. 
 A stationary state has been obtained.  Turbulence is sustained by outflows 
which  counteract gravity, delaying the collapse significantly.
\citet{wang2010} and \citet{fed2015} carried out  simulations in which several  
physical processes are progressively included, namely initial turbulence, magnetic field and outflows. 
Each of them reduces the star formation rate  by a factor of a few.
When all of them are included
 the star formation rate  is typically 10 times lower than when the protocluster is in freefall.

 The question as to whether turbulence may be sustained by protostellar outflows has been 
 investigated by 
\citet{cunningham2009} and \citet{carroll2009}. They reached the conclusion 
that in a turbulent medium, even without a magnetic field, the 
outflows  couple to the surrounding gas and trigger turbulence efficiently. 
They inferred an energy powerspectrum that is 
 stiffer than the usual  powerspectra found in large scale driven turbulence 
\citep[e.g.][]{kritsuk2007,hf2012}).
\citet{murray2018} have reached a somehow different conclusion as they find that the outflows 
have only a modest influence on the driving of turbulence. 

\citet{offner2017} and \citet{offner2018} have performed a series of low mass dense core collapse
and studied the influence of outflows on the collapsing core and in particular,  the efficiency of the 
driving of turbulence within the envelope of the core. They conclude that outflows can drive efficiently
turbulence in the envelope and that the efficiency of the driving increases with magnetic intensity.

\section{The role of magnetic field in ISM self-regulated models}
\label{section_ls}

Important efforts have also been undertaken to self-consistently simulate 
the interstellar medium within galaxies. Because modelling galaxies 
as a whole is very challenging in terms of scales, many models \citep[e.g.][]{deav2005,joung2006,kim2013,gent2013b}
consider a computational box of about 1 kpc sometimes called galactic box. Since the typical supernova
remnant radius is about 50 pc, this constitutes a good compromise between 
spatial resolution and molecular cloud statistics (though at the expense of 
solving the large galactic scales). The most recent models consider an external vertical 
gravitational field, which represents the gravity of stars and dark matter, 
follow the star formation (up to spatial scales of about 1-4 pc) and deliver 
stellar feedback (due to massive stars and essentially though not exclusively supernovae). 
This leads to a self-regulated ISM in which a turbulent cascade takes place. 
The energy is injected at the large and intermediate (around or above 100 pc) scales
 and decay at the small ones.

\subsection{Star formation rate and vertical equilibrium}
The importance of the spatial and temporal correlations between the supernova 
remnants and the star forming dense gas has been stressed by recent studies \citep{hi2014,gatto2015}. 
When the supernovae are  randomly placed in the,  the feedback they provide 
is inefficient and does not reduce the star formation 
rate appreciably. On the other hand, when the supernova explosions
correlate  with the dense gas, star formation rates in better agreement  with the 
observed values are inferred \citep{kim2013,hi2014}. 
 In these simulations, the thickness of the galactic disk is also compatible 
with the observed values while it is too thin  in simulations where the supernovae
are randomly placed. As recently stressed by \citet{girich2016} cosmic rays may change 
this conclusion.
Also these models  produce a realistic  multi-phase magnetized ISM
with densities and temperature that are reminiscent of the WNM and CNM.  
When a magnetic field of a few $\mu$G is initially present in the simulations, the   magnetic intensities
stay  compatible with the observed values. It has been found that magnetic field contributes 
to the galactic vertical equilibrium although its contribution is lower than the one of the turbulent dispersion
 and it has also been inferred that 
the star formation rate is somewhat reduced in the presence of a magnetic field 
by a factor that is on the order of 2 \citep{kim2015,iffrig2017,girich2018}. 
One important limit of these models is that the feedback is delivered immediately after the stars are formed,
 while supernovae 
arise  4-40 Myr after  their progenitor formation. Given that  the typical freefall time  of a typical star forming cloud
is only a few Myr. This is a significant effect that 
the most advanced models \citep{tigress2016,colling2018,girich2018} 
are now taking into account. It should be stressed however that 
in order  to treat the feedback injection properly  star formation and evolution should be treated self-consistently. 
In practice, this would  require to resolve   spatial scales that are much smaller 
than  what is currently possible for this type of modelling. 

\subsection{Turbulence and clumps} 
\citet{iffrig2017} have carried out a series  of 1024$^3$ simulations  which allow to infer the statistics
of turbulence and the properties of structures. In spite of 
the  stratification, the   powerspectra are broadly compatible with earlier 
works \citep[see e.g.][]{kritsuk2007} though the velocity powerspectrum is closer to the 
classical Kolmogorov exponent than the stiffer, almost Burgers like, values inferred in supersonic 
isothermal turbulence. This likely is a consequence of the magnetized, multi-phase structure
since the velocity dispersion is not much larger than the sound speed and Alfv\'en speed 
of the WNM.  
The ratio of the energies of the compressible modes and solenoidal ones 
depends on the  altitude. In the mid-plane, the compressible modes dominate 
while above a certain altitude, which varies with the magnetic intensity, the solenoidal ones 
dominate. The stronger the magnetic intensity, the lower is the altitude above which solenoidal modes 
dominate.
This conclusion is  different from the one of \citet{padoan2016} who found that 
the solenoidal modes always dominate. The discrepancy  is most certainly due
 to the absence of stratification in \citet{padoan2016}.
The dense clouds have been extracted from the simulations of \citet{iffrig2017}
and \citet{padoan2016} using  simple clump finders. 
Their statistical properties such as the mass spectra, the mass-size and the 
internal velocity dispersion-size relations are all reminiscent of the observed cloud properties 
\citep[e.g.][]{mivi2017} though \citet{iffrig2017} mentioned that the internal velocity dispersion
are possibly smaller than within observed clouds. This may indicate the need for other energy 
injection sources such as the one due to the large galactic scale gravitational instabilities \citep{krum2016}. 
The distribution of the mass-to-flux ratio, $\mu$, of the clouds has also been inferred \citep[see also][]{inoue2012}. 
It is broadly proportional to
the square-root of the cloud mass, which has been interpreted as the mass being proportional to the volume while 
the flux is proportional to the surface. 
 The value of $\mu$ also depends on the density threshold used to define the clouds. 
The lower the density threshold, the lower $\mu$. 

\subsection{A possible link between magnetic field and clump mass function} 
To understand the overall star formation rates in the Galaxy we have to know, not only the star formation rate in an 
individual cloud, but also the mass distribution of molecular clouds, which determines the total number of stars created in the Galaxy. 
It is actually difficult to accurately determine the mass function of molecular clouds in our Galaxy because of the line-of-sight contamination and limited knowledge on the distances to the clouds. 
Thanks to the development of observations,  the mass function of GMC  
 can now be determined  in nearby face-on galaxies such as M51 \citep{Colombo+2014}. 
For example \citet{Colombo+2014} reported that the exponent of the power-law slope of mass function varies depending on relative location to the spiral arm structure and the galactic center. 
Thus theoretical studies for the cloud properties may shed light on our understanding of the formation and destruction of molecular clouds. 
As mentioned in the previous sections, however, it is still difficult to perform direct numerical simulations of an ensemble of molecular clouds and study in details the small scale physics such as formation and destruction of molecular clouds. 
Earlier attempts to propose analytical models can be found in \citet{Kwan1979}, \citet{ScovilleHersh1979}, 
and \citet{Tomisaka1986} that formulated the so-called coagulation equation for molecular clouds. 
In these investigation the growth of clouds are, however, supposed to be driven by the cloud-cloud collision and
 missed the importance of gas accretion onto molecular clouds. 
The recent theoretical finding of the long timescale of molecular cloud formation \citep{inoue2009} and the importance of gradual growth process by accretion of dense HI gas \citep{inoue2012} stress the crucial need for accretion contribution in 
the coagulation equation \citep{Kobayashi+2017,Kobayashi+2018}. 

In this section, we present an analytical model that suggests a link between the magnetic field and  
the clump mass function because of the impact of the former onto the cloud formation time.
In Section 3.3 we have shown that the existence of magnetic field may possibly significantly increase 
the formation timescale of molecular clouds.  
Let's propose an estimate of the actual value of the cloud formation timescale in our Galaxy. 
The radius of a supernova remnant (SNR) can be on the order of $100$pc after the expansion over the typical age $\sim$ 1 Myr. 
We may assume that the creation rate of SNRs in our Galaxy is $10^{-2} {\rm yr}^{-1}$ 
Thus, the volume occupied by SNRs can be calculated as $100^{3} \times 10^{-2} {\rm yr}^{-1} \times 1{\rm Myr} = 10^{10} {\rm pc}^3$. 
This value is roughly the same as the volume of Galactic thin disk ($10 {\rm kpc}^2 \times 100 {\rm pc}$) where molecular clouds reside. 
This means that ISM in Galactic thin disk is swept up by SNR once per 1 Myr \citep{McKeeOstriker1977}.
If we ignore the magnetic field, the molecular cloud can be simply created by a single, may be a few
 compressions of warm neutral medium by the propagation of a shock wave. 
As shown in Section 3.3, however, molecular clouds could be created after several compression (up to $10$) 
 and thus, the actual timescale of cloud formation should be several Myr. 

To infer the clump mass spectrum, we can adopt coarse graining of short-timescale ($\sim$ a few Myr) events of the growth 
and destruction of clouds, and describe the long timescale evolution by the continuity equation of molecular clouds 
in mass space \citep{Kobayashi+2017}
\begin{equation}
  \PD{N}{t} + \PD{}{M} \left( N \frac{dM}{dt} \right)
               =  - \frac{N}{T_{\rm d}} + \left( \frac{dN}{dt} \right)_{\rm coll},  
\end{equation}
where 
$N(dM/dt)$ denotes the flux of mass function in mass space, 
$T_{\rm d}$ is the cloud disruption timescale, 
$dM/dt$ describes the growth rate of the molecular cloud, 
and the last term accounts for the growth due to cloud-cloud collision. 
If the contribution from cloud-cloud collisions is negligible \citep{Kobayashi+2017,Kobayashi+2018} and the mass growth can be approximated by 
$dM/dt = M/T_{\rm f}$ with the growth timescale $T_{\rm f}$, 
a steady state solution of the above equation is 
$N(M) = {M}^{-\alpha}$, where $\alpha = 1 + T_{\rm f}/T_{\rm d}$ \citep{inutsuka2015}. 
In a gas rich environment such as a spiral arm of a disk galaxy, we expect $T_* \sim T_{\rm f}$, and thus, $T_{\rm f} \lesssim T_{\rm d}$, which corresponds to $ 1 < \alpha \lesssim 2 $.
For example, $T_{\rm f}=10$Myr corresponds to $\alpha \approx 1.7$, which agrees nicely with observations \citep{Solomon+1987,Kramer+1998,Heyer+2001,RomanDuval+2010}. 
However, in a region with very limited amount of gaseous material, $T_{\rm f}$ is expected to be large and possibly even larger than $T_{\rm d}=T_* + 4$Myr, which produces $\alpha > 2$.
This may explain the observations in M33 \citep{Gratier+2012} and in M51 \citep{Colombo+2014}. 
The more detailed description of the molecular cloud mass function can be found in \cite{Kobayashi+2017,Kobayashi+2018} where the effect of cloud-cloud collisions is explicitly taken into account. 

Note that the effects of magnetic field that slows down the cloud formation are taken into account in the above analysis as a large value of the cloud formation timescale ($T_{\rm f} > 1$ Myr). 
If we ignore the effect of magnetic field and simply choose the dynamical compression rate of ISM as the value of the cloud formation timescale $T_{\rm f}$=1Myr, the powerlaw exponent of the mass function of molecular cloud would be too small ($\alpha \sim$ 1), which is in stark contrast to the observed values. 
Therefore we may conclude that magnetic field is playing an important role in the mass distribution of molecular clouds in our Galaxy.

\section{Conclusions}
This review is dedicated to the role that magnetic field 
may have in the formation and evolution of molecular clouds. 
Significant progress has been accomplished in the last years
in our understanding of the molecular cloud in particular and 
star formation process in general. 
We have a better, although still incomplete, knowledge of the 
structures, filaments, cores, clumps, clusters, formation mechanisms. 

Most likely these gaseous structures are all the product of magnetized turbulence 
interacting with gravity. Given the values of the 
magnetic intensities that have been measured, numerical 
simulations seem to indicate that the number of objects 
that form at all scales, from clumps to stars, is likely 
reduced by a factor of a few  due to the action of the magnetic field. 
Accordingly their masses tend to be also a few times larger 
than what it would be with  pure hydrodynamics. 
The shapes of the clouds
are also strongly affected by magnetic field, which
tends to create filamentary structures as well as 
clouds that have flattened along the magnetic field lines that permeate them. 
More generally the whole dynamics of the ISM is significantly 
modified and cannot be accurately interpreted without taking 
magnetic fields into account.

While it is now almost certain that magnetic fields 
do not regulate the star formation process by reducing 
the star formation rate drastically, as proposed 
three decades ago, 
it is likely the case that magnetic fields 
contribute to reduce it  by a
factor of a few.  
Moreover since it has been found by various groups that 
magnetic field tends to reduce the fragmentation 
and to produce  stars with larger mass, 
another possible consequence of magnetic field is to 
enhance stellar feedback and therefore to reduce
the star formation rate and efficiency in 
molecular clouds. This latter aspect remains however to be confirmed as 
numerical simulations are not able now to cover the necessary range of scales.
Finally we stress that magnetic field is likely to have drastic consequences on the formation 
of protoplanetary disks through magnetic braking by reducing and even possibly controlling 
their size.

\section*{Funding}
This work is supported by Grant-in-aids from the Ministry of Education, Culture, Sports, Science, and Technology (MEXT) of Japan (15K05039 and 16H02160). 

\section*{Acknowledgments}
We thank the two referees for constructive remarks and critical reading of the manuscript.
We thank various contributions from 
 Philippe Andr\'e, Edouard Audit, Doris Arzoumanian, Gilles Chabrier, 
Benoit Commer{\c c}on, Edith Falgarone, S\'ebastien Fromang, Samuel Geen, 
 Olivier Iffrig, Tsuyoshi Inoue, Kazunari Iwasaki, Hiroshi Koyama, Yueh-Ning Lee, Anaelle Maury, 
Evangelia Ntormousi, Juan-Diego Soler, Romain Teyssier,  Valeska Valdivia.






\end{document}